\documentclass[aps,prd,showkeys,showpacs,twocolumn,nofootinbib,nobibnotes,
superscriptaddress,preprintnumbers,notitlepage,longbibliography]{revtex4-1}

\usepackage[utf8]{inputenc} 
\usepackage[top=2.8cm, bottom=2.8cm, left=2cm, right=2cm]{geometry}
\usepackage[T1]{fontenc}     
\usepackage{bm}
\usepackage{amsmath,amssymb,lmodern} 
\usepackage{sidecap}
\usepackage[caption=false]{subfig}
\usepackage{graphicx}
\usepackage{tabularx}
\usepackage{enumerate} 
\usepackage[caption=false]{subfig}
\usepackage{natbib}
\usepackage{bm}
\usepackage{makecell}
\usepackage{hyperref}\hypersetup{
    colorlinks,%
    citecolor=blue,%
    filecolor=blue,%
    linkcolor=blue,%
    urlcolor=blue
}
\usepackage{color}
\usepackage{cleveref}
\allowdisplaybreaks[1]

\newcommand{\be}{\begin{equation}}
\newcommand{\ee}{\end{equation}}
\newcommand{\bea}{\begin{eqnarray}}
\newcommand{\eea}{\end{eqnarray}}
\newcommand{\ben}{\begin{enumerate}}
\newcommand{\een}{\end{enumerate}}

\begin{document}

%\preprint{PI/UAN-2020-672FT}

\title{Dynamical analysis of cosmological models with non-Abelian gauge vector fields
%Inflation and dark energy from non-Abelian gauge vector fields
}

\author{Alejandro Guarnizo}
\email{alejandro.guarnizo@correounivalle.edu.co}
\affiliation{Departamento de F\'isica, Universidad Antonio Nari\~no, \\ Cra 3 Este No. 47A-15, Bogot\'a DC, Colombia}
\affiliation{Departamento de F\'isica, Universidad del Valle,\\
Ciudad Universitaria Mel\'endez, Santiago de Cali 760032, Colombia}

\author{J. Bayron Orjuela-Quintana}
\email{john.orjuela@correounivalle.edu.co}
\affiliation{Departamento de F\'isica, Universidad del Valle,\\
Ciudad Universitaria Mel\'endez, Santiago de Cali 760032, Colombia}

\author{C\'esar A. Valenzuela-Toledo}
\email{cesar.valenzuela@correounivalle.edu.co}
\affiliation{Departamento de F\'isica, Universidad del Valle,\\
Ciudad Universitaria Mel\'endez, Santiago de Cali 760032, Colombia}

%\date{\today}

%%%%%%%%%%%%%%%%%%%%%%%%%%%%%%%%%%%%

\begin{abstract}

In this paper we study some models where non-Abelian gauge vector fields endowed with a SU(2) group representation are the unique source of inflation and dark energy. These models were first introduced under the name of gaugeflation and gaugessence, respectively. Although several realizations of these models have been discussed, not all available parameters and initial conditions are known. In this work, we use a dynamical system approach to find the full parameter space of the massive version of each model. In particular, we found that the inclusion of the mass term increases the length of the inflationary period. Additionally, the mass term implies new behaviors for the equation of state of dark energy allowing to distinguish this  from other prototypical models of accelerated expansion. We show that an axially symmetric gauge field can support an anisotropic accelerated expansion within the observational bounds.

\end{abstract}



\maketitle

%%%%%%%%%%%%%%%%%%%%%%%%%%%%%%%%%%%%

\section{Introduction}

Inflation, an early accelerated expansion of the Universe, is arguably the most compelling theory to overcome the classical problems of the hot big bang cosmology while providing the seeds for large scale structure formation \cite{liddle_lyth_2000}. The simplest inflationary models are based in the dynamics of a scalar field, the inflaton. Although the inflaton is generally favored by data \cite{Akrami:2018odb}, some particular characteristics observed in the cosmic microwave background (CMB), known as CMB anomalies, suggest that modifications to this paradigm might be needed.  For instance, careful analysis of the CMB indicates an asymmetry in the dipolar power spectrum on large angular scales \cite{Groeneboom:2009cb, Kim:2013gka, Ramazanov:2013wea}. This particular problem, which has the largest statistical significance among the different anomalies involving a preferred direction in the CMB sky\footnote{For a review of the cosmic anomalies see Refs. \cite{Schwarz:2015cma, Perivolaropoulos:2014lua}.} \cite{Ade:2015hxq}, motivates the study of alternative models based on other types of fields. 

Apart from the early inflationary period, it is an observational fact that the current Universe is also undergoing an accelerated expansion \cite{Perlmutter:1998,Riess:1998}. In this case, the simplest explanation is provided by the cosmological constant $\Lambda$ \cite{Amendola:2015ksp}. However,  despite its success,  this scenario has some troubles when it is compared with observations \cite{Martin:2012bt}. One of these difficulties is related to the fundamental nature of $\Lambda$ as the vacuum energy density of the Universe. This identification results in a huge discrepancy between the value predicted by the theory and the value obtained from observations. This disagreement (around $120$ orders of magnitude) is usually referred to as the cosmological constant problem \cite{Weinberg:1988cp,Amendola:2015ksp}.  Another issue is the so-called $H_0$ tension \cite{Riess:2016jrr, Riess:2019cxk}, which states that the current value of the Hubble parameter calculated from CMB data does not agree with the value computed from local measurements.  By introducing a dynamical equation of state for the source driving the late-time accelerated expansion, this problem could be addressed \cite{Guo:2018ans}. These problems suggest that new dynamical degrees of freedom must be considered and the most popular models are the so-called quintessence models, which are based, again, on scalar fields \cite{Copeland:2006wr,Tsujikawa:2010sc,Yoo:2012ug}.

As mentioned above, the most popular models to account for the inflationary period and the current accelerated expansion of the Universe are based on single scalar fields. Despite the successes of these theories, other interesting alternatives, built with different types of fields, have been also explored. In this direction, models that include vector fields (see Refs. \cite{Dimastrogiovanni_2010,Maleknejad:2012fw,Soda:2012zm,Heisenberg:2018vsk}  for reviews on the subject), higher spin fields \cite{Shiraishi:2013vja,Bartolo:2017sbu,MoradinezhadDizgah:2018ssw,Franciolini:2018eno,MoradinezhadDizgah:2018pfo,Bordin:2019tyb}, $p$-forms \cite{Koivisto:2009sd,Koivisto:2009fb,Koivisto:2012xm,Mulryne:2012ax,Kumar:2016tdn,Almeida:2018fwe,Almeida:2019xzt,Almeida:2019iqp,Guarnizo:2019mwf,Almeida:2019xzt,Almeida:2020lsn}, among others, have called the attention within the past years because of their richer phenomenology and since many of their cosmological consequences have not been fully explored so far. Between these proposals, non-Abelian gauge fields have recently attracted a lot of attention since they could link cosmology with the phenomenology of particle physics \cite{Maleknejad:2012fw}. The cosmological dynamics of these fields have been extensively discussed in the literature  \cite{Emoto:2002fb,Hosotani:2002nq,Maleknejad:2011jw,Maleknejad:2011sq,Maleknejad:2011jr,Murata:2011wv,Noorbala:2012fh,Maleknejad:2012dt,Adshead:2012kp,Maleknejad:2013npa,Rinaldi:2014yta,Rinaldi:2015iza,Nieto:2016gnp,Adshead:2016omu,Adshead:2017hnc,Rodriguez:2017wkg,Mehrabi:2017xga,Alvarez:2019ues, Gomez:2019tbj, Wolfson:2020fqz,Gomez:2020sfz,Orjuela-Quintana:2020klr, SheikhJabbari:2012qf, Landim:2016dxh}. For instance, the early and the late-time accelerated expansions can be uniquely explained by non-Abelian gauge vector fields. These models are known as ``gaugeflation"  \cite{Maleknejad:2011jw, Maleknejad:2011sq} and ``gaugessence" \cite{Mehrabi:2017xga}, respectively. Regarding the early accelerated expansion, gaugeflation was ruled out as a valid inflationary model, since the relation between the scalar spectral tilt and the tensor-to-scalar ratio does not get in the region allowed by Planck 2013 \cite{Planck:2013jfk, Namba:2013kia}. Nevertheless, the introduction of a mass term in the theory can alleviate the problem \cite{Nieto:2016gnp, Adshead:2017hnc}. Some of the background inflationary trajectories of this proposal were studied in Refs. \cite{Nieto:2016gnp, Adshead:2017hnc}, where it was concluded that, in particular, the addition of the mass term does not affect the existence of an inflationary period but it does reduce its length. Here, by using a dynamical system approach, we find the full available parameter space of this model and show that the inclusion of the mass term actually increases the length of inflation, instead of reducing it. We also remark that this particular result could yield to modifications at the linear perturbation level as worked out in Ref. \cite{Adshead:2017hnc}. On the other side, in Ref.\cite{Mehrabi:2017xga} the gaugessence model was proposed as a possible explanation to the late-time accelerated expansion. It was shown that for several sets of initial conditions and parameters there exists a period of accelerated expansion. In this work, we generalize this model by considering the effect of a mass term for the gauge vector field in the late-time cosmological evolution. We show in particular that accelerated expansion is an ``effective" attractor. We also study the equation of state of dark energy, showing different behaviors to those found in Ref. \cite{Mehrabi:2017xga} that allow distinguishing this model from other dark energy proposals. We also address some differences between this model and the usual quintessence models. 

Besides the CMB anomalies mentioned before, several observational indications are suggesting that the present Universe is undergoing an anisotropic accelerated expansion \cite{Campanelli:2010zx, Antoniou:2010gw, Zhao:2016fas, Salehi:2016sta, Amirhashchi:2018nxl}. Since scalar fields cannot pick a preferred direction in spacetime, it is natural to consider other types of dynamical fields. Regarding gauge vector fields, in Ref. \cite{Maleknejad:2011jr} was shown that an  axially symmetrical massless gauge vector field isotropizes during the inflationary expansion, and thus any initial anisotropy is quickly dilute. However, this scenario has not been studied in the frame of the late time accelerated expansion where other conclusions or interesting features can be reached. Here we consider the dynamics of an axially symmetrical gauge field in a homogeneous but anisotropic background and numerically investigate the possibility to get a non-negligible contribution to the current spatial shear.

The paper is organized as follows. In Sec. \ref{sec:inflation} we study the inflationary dynamics of a particular model for a massive non-Abelian gauge vector field, and its massless version, at a background level. We present a dynamical system analysis which allows us to constrain the full parameter space of the theory and thus to extend the results in previous works. In particular, we show that the inclusion of the mass term increases the length of inflation rather than reduce it. In Sec. \ref{dark energy} we study the late-time cosmological behaviour of these gauge fields as dark energy components, generalizing previous works by including a mass term to the dynamics. By analyzing the equation of state of dark energy, we show that new behaviors are found. Next, we investigate the possibility to get non-negligible contributions to the spatial shear today by considering anisotropic solutions in the massless case. Finally, our conclusions are presented in Sec. \ref{conclusions}.

%%%%%%%%%%%%%%%%%%%%%%%%%%%%%%%%%%%%

\section{Inflation from gauge fields} \label{sec:inflation}

Let us consider the following action for a massive SU(2) gauge vector field
\begin{multline} \label{gauge action}
S = \int  \text{d}^4 x  \sqrt{- 	\tilde{g}} \Biggl[\frac{m_{\text{P}}^2}{2} R - \frac{1}{4}F_{\,\, \mu\nu}^a F_a^{\,\, \mu\nu} 
\\
+ \frac{\kappa}{96}\left(\tilde{F}^{\,\, \mu\nu}_a F^a_{\,\, \mu\nu}\right)^2 - \frac{1}{2} m^2_a A_{\,\, \mu}^a A^{\,\, \mu}_a \Biggr] \,,
\end{multline}
where $g_{\mu\nu}$ is the metric of the spacetime, $\tilde{g}$ its determinant, $m_\text{P}$ is the reduced Planck mass, $R$ is the Ricci scalar,  $\kappa$ is a positive-definite constant with dimensions $[\, m_\text{P}^{- 4}\,]$, $F^a_{\,\, \mu\nu}$ is the field strength tensor\footnote{Greek indices run from 0 to 3 and denote space-time components, and Latin indices run from 1 to 3 and denote SU(2) gauge components.}
\begin{equation}
F^a_{\,\, \mu\nu} \equiv \partial_\mu A_{\,\, \nu}^a - \partial_\nu A_{\,\, \mu}^a + g \, \varepsilon^a_{\,\, b c} A^b_{\,\, \mu} A_{\,\, \nu}^c \,,
\end{equation}
of a non-Abelian gauge vector field $A^a_{\,\, \mu}$ with mass $m_a$, $g$ is the SU(2) coupling constant and $\varepsilon^a_{\,\,b c}$ being the Levi-Civita symbol. The dual of the strength tensor is defined as usual by
\begin{equation}
\tilde{F}^{\,\, \mu\nu}_a \equiv \frac{1}{2} \varepsilon^{\mu\nu\rho\sigma} F^a_{ \,\, \rho\sigma} \,, 
\end{equation}
with $\varepsilon^{\mu\nu\rho\sigma}$ denoting the completely antisymmetric tensor. This action was first studied in Ref. \cite{Nieto:2016gnp}, in the context of inflation, as a modification of the original model in Ref. \cite{Maleknejad:2011sq} where massless fields were considered.

Observations show that both the early and the late Universes are highly homogeneous, isotropic and spatially flat \cite{Ade:2015xua}. It allows us to describe it, at the background level, by the Friedmann-Lema\^itre-Robertson-Walker (FLRW) metric
\begin{equation} \label{flrw metric}
\text{d} s^2 = - \text{d} t^2 + a^2(t) \delta_{i j} \text{d} x^i \text{d} x^j \,,
\end{equation}
where $a(t)$ is the scale factor of the expansion, $t$ the cosmic time, and $x^i$ the Cartesian coordinates. An ansatz for the massive gauge vector field consistent with the symmetries of this spacetime is \cite{Bento:1992wy}
\begin{equation} \label{ansatz}
A^a_{\,\, 0} \equiv 0 \,, \quad  A^a_{\,\, i} \equiv a(t) \psi (t) \delta^a_{\,\, i} \,,
\end{equation}
where $\psi(t)$ is a scalar field. As a shorthand notation, we define $\phi(t) \equiv a(t) \psi (t)$. Isotropy also requires that the gauge fields have identical masses, i.e. $m_1 = m_2 = m_3 = m$. 

Varying the action in Eq. (\ref{gauge action}) with respect to the metric $g^{\mu\nu}$ we obtain the energy tensor 
\begin{multline} \label{energy tensor}
T_{\mu\nu} = F^a_{\,\, \mu\rho} F^a_{\,\, \nu\sigma} g^{\sigma\rho} + m^2 A_{\,\, \mu}^a A_{\,\, \nu}^a - g_{\mu\nu} \Biggl[ \frac{1}{4} F^{\,\, \rho\sigma}_a F^a_{\,\, \rho\sigma}  \\
+ \frac{\kappa}{96}\left(\tilde{F}^{\,\, \rho\sigma}_a F^a_{\,\, \rho\sigma}\right)^2 + \frac{1}{2} m^2 A^{\,\,\lambda}_a A_{\,\, \lambda}^a \Biggr] \, .
\end{multline}
Employing the ansatz in Eq. (\ref{ansatz}), the corresponding energy density $\rho$ and pressure $p$ can be written in terms of three contributions: the first one coming from the Yang-Mills term, the second one arising from the $\kappa$-term (i.e. the $(F \tilde{F})^2$ term) and the last one from the mass term
\begin{equation*} 
\rho = \rho_{\text{YM}} + \rho_\kappa + \rho_A \, , \, 
\end{equation*}
\begin{equation} \label{rho p}
p = \frac{1}{3}\rho_{\text{YM}} - \rho_\kappa - \frac{1}{3} \rho_A\, , 
\end{equation}
where
\begin{equation*} 
\rho_{\text{YM}} \equiv  \frac{3}{2} \left(\frac{\dot{\phi}^2}{a^2} + \frac{g^2 \phi^4}{a^4} \right) \, , \quad \rho_\kappa \equiv \frac{3}{2} \kappa g^2 \frac{\phi^4 \dot{\phi}^2}{ a^6} \,,
\end{equation*}
\begin{equation} \label{densities}
\rho_A \equiv \frac{3}{2} \frac{m^2 \phi^2}{a^2} \,.
\end{equation}

The Friedmann equations read
\begin{equation} \label{inf H2}
3 m_\text{P}^2 H^2 = \frac{3}{2} \left[ \frac{\dot{\phi}^2}{a^2} +  \frac{g^2 \phi^4}{a^4} + \kappa g^2 \frac{\phi^4 \dot{\phi}^2}{a^6} +  \frac{m^2 \phi^2}{a^2} \right] \,, 
\end{equation}
\begin{equation} \label{inf Hdot}
2 m_\text{P}^2 \dot{H} = - 2 \left[ \frac{\dot{\phi}^2}{a^2} + \frac{g^2 \phi^4}{a^4} + \frac{1}{2} \frac{m^2 \phi^2}{a^2} \right] \,,
\end{equation}
$H \equiv \dot{a} / a$ being the Hubble parameter. Varying the action in Eq. (\ref{gauge action}) with respect to $A^a_{\,\, \nu}$ we get
\begin{multline} \label{field variation}
0 = \nabla_\mu \left[ F^{\,\, \mu\nu}_a - \frac{\kappa}{12} \left( \tilde{F}^{\,\, \rho\sigma}_d F^d_{\,\, \rho\sigma} \right) \tilde{F}^{\,\, \mu\nu}_a \right]  - m^2 A^{\,\, \nu}_a \\
 - g \varepsilon^{\,\, ab}_c A_{\,\, \mu}^c \left[ F^{\,\, \mu\nu}_b - \frac{\kappa}{12} \left( \tilde{F}^{\,\, \rho\sigma}_d F^d_{\,\, \rho\sigma} \right) \tilde{F}^{\,\, \mu\nu}_b \right] \,.
\end{multline}
By using Eq. (\ref{ansatz}), the only nontrivial equation of motion for the gauge fields is
\begin{multline} \label{field eom}
0 = \frac{\ddot{\phi}}{a} \left( 1 + \kappa g^2 \frac{\phi^4}{a^4}\right) + \frac{H \dot{\phi}}{a}\left(1 - 3\kappa g^2 \frac{\phi^4}{a^4}\right)  \\
 + \frac{2 g^2 \phi^3}{a^3}\left(1 + \kappa \frac{\dot{\phi}^2}{a^2}\right) + m^2 \frac{\phi}{a}\,.
\end{multline} 
Equations (\ref{inf H2}), (\ref{inf Hdot}), and (\ref{field eom}) give the dynamics of the inflationary phase. It is more convenient to recast a set of nonlinear equations in terms of dimensionless expansion variables \cite{Wainwright2009}. In our case we choose
\begin{equation}
x \equiv \frac{1}{\sqrt{2} m_\text{P}} \frac{\dot{\phi}}{aH} \,, \quad  y \equiv \frac{1}{\sqrt{2} m_\text{P}}\frac{g \, \phi^2}{a^2 H} \, , 
\end{equation}
\begin{equation}  \label{inf variables}
w \equiv \frac{1}{\sqrt{2} m_\text{P}}\frac{m \phi}{a H}\,, \quad  z \equiv \frac{1}{\sqrt{2} m_\text{P}}\frac{\phi}{a} \,.
\end{equation}
The Friedmann equation in Eq. (\ref{inf H2}) becomes the constraint
\begin{equation} \label{inf constraint}
1 = x^2 + y^2 + w^2 + 4 \alpha x^2 z^4 \,, 
\end{equation}
from which we can write the variable $y$ as a function of the other variables and a dimensionless parameter defined as $\alpha \equiv m_{\text{P}}^4 \, \kappa g^2$.
By changing the cosmic time $t$ with the number of $e$-folds $N$ defined by $\text{d} N \equiv H \text{d} t$, and taking into account Eqs. (\ref{inf Hdot}), (\ref{field eom}) and the constraint in Eq. (\ref{inf constraint}), we can write the evolution equation for each independent variable as follows:
\begin{align} 
x' &= x (\epsilon - 1) \nonumber \\
 &- \left(1 + 4 \alpha z^4\right)^{-1} \left[ \frac{2 }{z}(1 - x^2) + x \left(1 - 12 \alpha z^4 \right) - \frac{w^2}{z} \right] \,, \label{inf x eq} \\
w' & = w \left( \frac{x}{z} + \epsilon - 1 \right) \,, \label{inf w eq} \\
z' &= x - z \,\,, \label{inf z eq}
\end{align}
where a prime denotes derivative with respect to $N$, and
\begin{equation} \label{inf slowroll}
\epsilon \equiv - \frac{\dot{H}}{H^2} = 2 - 8 \alpha x^2 z^4 - w^2 \,,
\end{equation}
is the slow-roll parameter. 

It is possible to calculate the expected number of $e$-folds of inflation at first order in the slow-roll approximation. Here we show the result for later use \cite{Nieto:2016gnp}
\begin{equation} \label{inf efolds}
N \approx \frac{1 + \gamma_i + \omega_i/2}{2 \epsilon_i} \,\, \text{ln} \left( \frac{1 + \gamma_i + \omega_i/2}{\gamma_i + \omega_i/2} \right) \,,
\end{equation}
where 
\begin{equation} \label{inf gamma}
\gamma \equiv \frac{g^2 \psi^2}{H^2} = \frac{y^2}{z^2} \,,
\end{equation}
\begin{equation} \label{inf omega}
\omega \equiv \frac{m^2}{H^2} = \frac{w^2}{z^2} \,,
\end{equation}
and the subindex $i$ indicates some initial time before the end of inflation.

For completeness, in the following subsection we first study the case where the gauge fields are massless.

%%%%%%%%%%%%%%%%%%%%%%%%%%%%%%%%%%%%

\subsection{Massless case}

This model was first introduced in Ref. \cite{Maleknejad:2011sq} as ``gaugeflation". Although several inflationary trajectories were analyzed in Ref. \cite{Maleknejad:2011sq}, an exploration of the full available parameter space of the theory has not been performed yet and thus only a few particular valid trajectories are known. In the following, we find the parameter window where slow-roll dynamics take place.

The massless case is characterized by $m = 0$ or equivalently $w = 0$. Equation (\ref{inf w eq}) is trivially satisfied and the dynamical system is reduced to Eqs. (\ref{inf x eq}) and (\ref{inf z eq}). The autonomous set has just one physically acceptable fixed point  given by\footnote{The fixed point is found upon regularization due to possible singularities in $z = 0$, and it is physically acceptable because the variables take real values and the Friedmann constraint (\ref{inf constraint}) is satisfied.}
\begin{equation} \label{inf fixed point}
x = z = \sqrt{\frac{\sqrt[3]{\alpha} f_{\alpha}^2 - \sqrt[3]{3}}{2 f_{\alpha} \sqrt[3]{9\alpha^2} }} \,,
\end{equation}
where
\begin{equation} \label{GF falpha}
 f_{\alpha} \equiv  \left( 9 + \sqrt{81 + \frac{3}{\alpha}} \right)^{1/3} \, .
\end{equation}
This point exists for all positive $\alpha$.

The slow-roll parameter in the fixed point is given by
\begin{equation} \label{GF slowroll}
\epsilon = 2 - 8\alpha z^6 = 2 - \frac{\left( f_\alpha^2 - \sqrt[3]{3}/\alpha^3 \right)^3}{9 f_\alpha^3} \,.
\end{equation}
Now, assuming the reasonably upper bound for enough slow-roll phase \cite{Karciauskas:2016pxn}
\begin{equation}
\epsilon < 10^{-2} \,,
\end{equation}
we find that 
\begin{equation} \label{GF alpha}
\alpha > 1.99 \times 10^{6} \,,
\end{equation} 
hence, in the massless case, the only relevant parameter of the system has to be very large in order to inflation has a correct length.

\begin{figure}[!t]
\centering
\includegraphics[width=\linewidth]{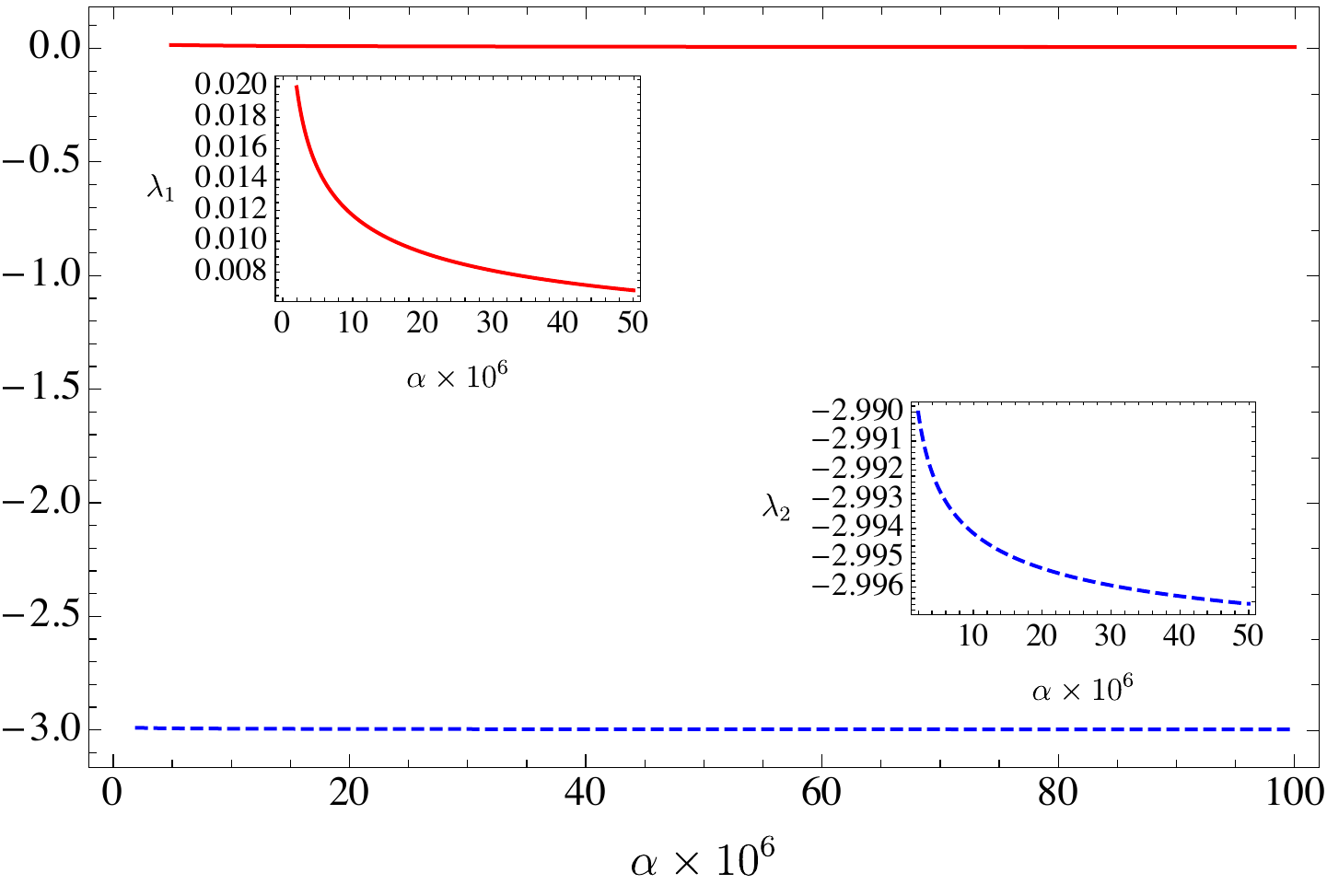}
\caption{Eigenvalues of the Jacobian matrix. The fixed point is a saddle since $\lambda_1 > 0 $ while $\lambda_2 < 0 $ in the region $\alpha > 2 \times 10^6$ where slow-roll solutions exist.}
\label{GF eigenvalues}
\end{figure}

The linear stability analysis shows that the fixed point is hyperbolic, i.e. the Jacobian matrix obtained from the dynamical system does not have any vanishing eigenvalue upon evaluation in the fixed point. In  Fig. \ref{GF eigenvalues} we plot the two eigenvalues as functions of the parameter $\alpha$ in the region where slow-roll solutions exist. We can see that one eigenvalue is positive, $\lambda_1 > 0$, and the other one is negative, $\lambda_2 < 0$, therefore the fixed point is a saddle in the space $(x, z)$. Moreover, we have corroborated that $\lambda_1 \rightarrow 0$ and $\lambda_2 \rightarrow - 3$ when $\alpha \rightarrow \infty$. This means that gaugeflation naturally agrees with the fact that inflation is a transient phase of the Universe. Inflation as a saddle fixed point has been addressed in other works (e.g. see Refs. \cite{Rodriguez:2015rua, Oikonomou:2017ppp}). Therefore, the dynamical analysis shows that the only fixed point of the system corresponds to a slow-roll inflationary solution for large $\alpha$.

Now, we will proceed to present a numerical solution to see the behaviour of the system under specific initial conditions. In Ref. \cite{Maleknejad:2011sq}, each inflationary trajectory is specified by a set of four values\footnote{Here, the subscript $i$ means that the corresponding quantity is evaluated some time before the end of inflation.} $(\psi_i, \dot{\psi}_i; \kappa, g)$. However, our dynamical analysis reveals that there is a particular point in the physical phase space $(x, z)$ which can be fully specified only by the parameter $\alpha$, being very useful for fixing the initial conditions. As an example of this particular slow-roll trajectory in the massless case ($\omega = 0$), we choose
\begin{equation}
\alpha = 2 \times 10^6 \, , 
\end{equation}
such that the variables in the fixed point in Eq. (\ref{inf fixed point}) are approximately 
\begin{equation}
x_i = z_i \approx 0.0706517 \,.
\end{equation}
The function $\gamma$ in Eq. (\ref{inf gamma}) and the slow-roll parameter in Eq. (\ref{GF slowroll}) take the values
\begin{equation}
\gamma_i \approx  4.8 \times 10^{-4} \,, \quad \epsilon_i \approx 9.9 \times 10^{-3} \,,
\end{equation}
which in Eq. \ref{inf efolds} translates into
\begin{equation} \label{GF 383 efolds}
N \approx 383 \,.
\end{equation}

\begin{figure}[t!]
%\subfloat[]
{
\centering
\includegraphics[width=\columnwidth]{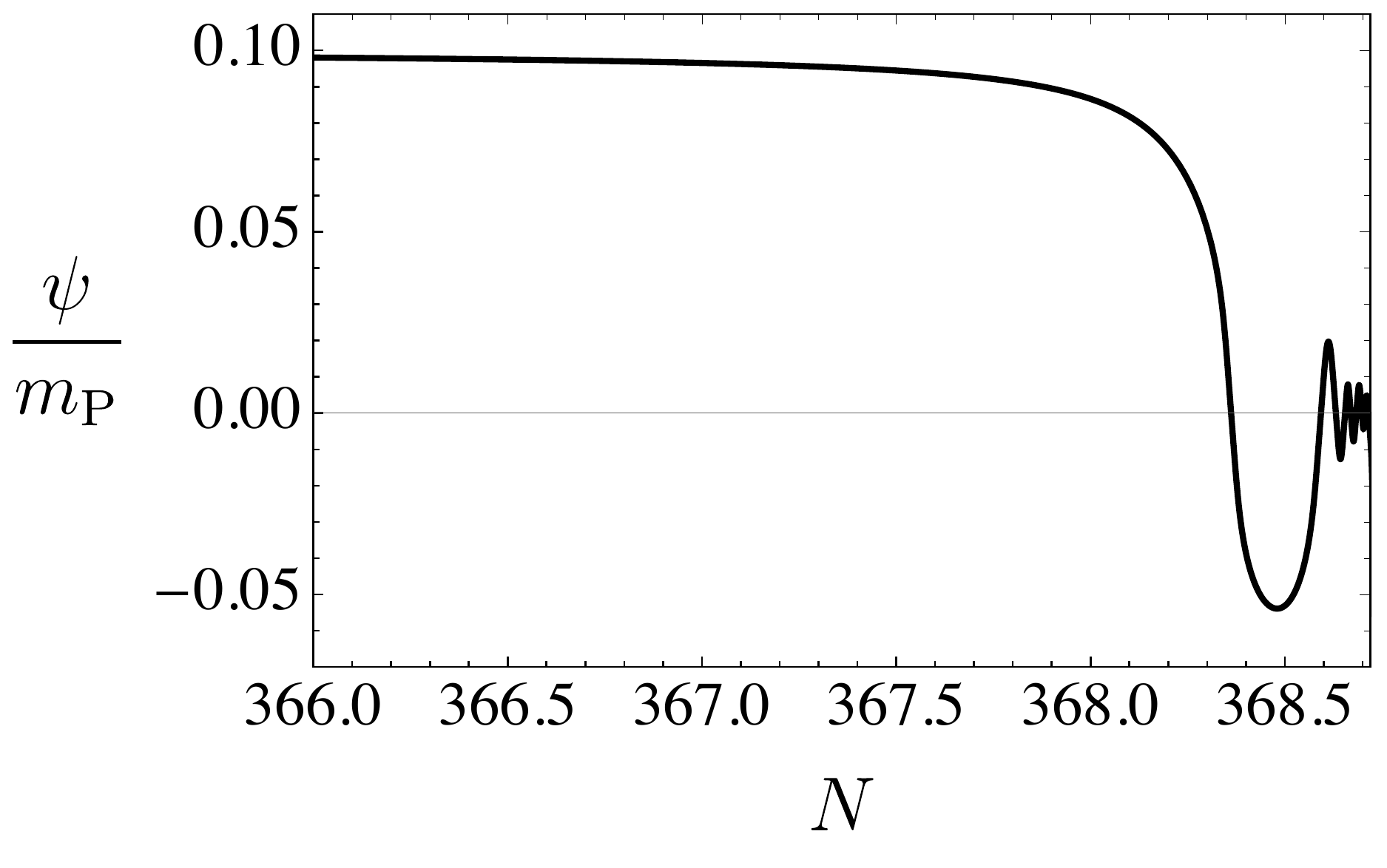}
%\label{GF field}
}
\vspace{0.05cm}
%\subfloat[]
{
  \centering
  \includegraphics[width=0.92\columnwidth]{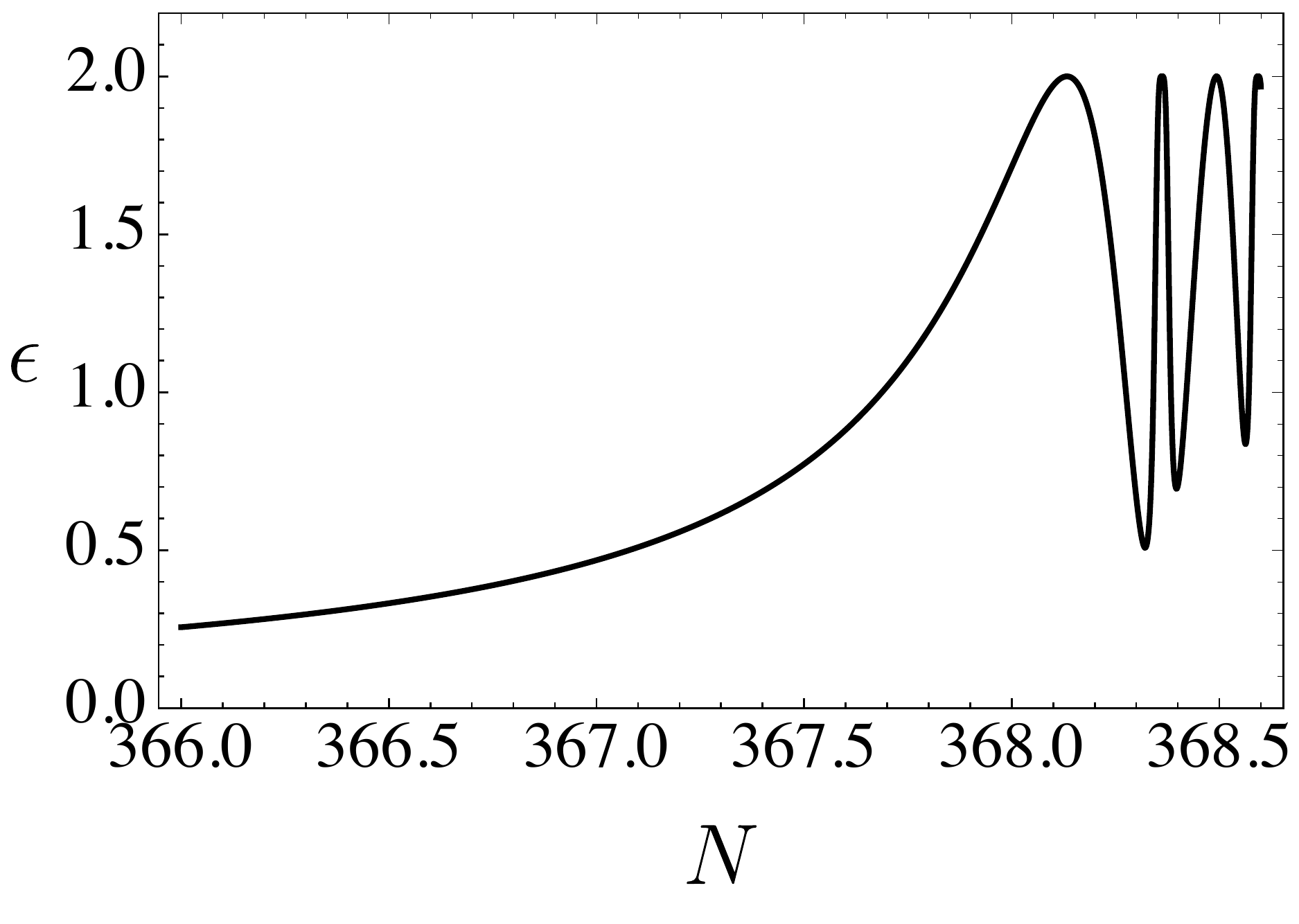}
%  \label{GF epsilon}
}
\caption{Evolution of the gauge field (upper) and the slow-roll parameter (lower). The field $\psi = \phi / a$ and the slow-roll parameter $\epsilon$ are nearly constant during inflation. The field suddenly decays while $\epsilon$ reaches its upper bound near to $N = 368$ around the value predicted in Eq. (\ref{GF 383 efolds}). The disagreement between these values is due to numerical approximation.}
\label{inf massless}
\end{figure}

As it can be seen in the upper part of Fig. \ref{inf massless}, for the chosen initial conditions, the gauge field is nearly constant during inflation, then it suddenly falls off around $N = 368$ and starts to oscillate, showing good matching with the analytical results. As explained in Ref. \cite{Maleknejad:2011sq}, these oscillations are presented because $\rho_{\text{YM}}$ becomes the dominant term at the end of inflation, and thus the system behaves as a quartic chaotic-like inflation theory. Support for this interpretation is given in the lower part of Fig. \ref{inf massless} where we see that the slow-roll parameter is small during inflation and it grows around $N = 368$, oscillating and reaching its upper limit equal to 2, which represents a ``dark radiation" domination epoch\footnote{By dark radiation we mean the ``radiation" associated to the Yang-Mills term.}. This is due to $\epsilon$ can be written as
\begin{equation} 
\epsilon = 2 \, \frac{\rho_{\text{YM}}}{\rho_{\text{YM}} + \rho_{\kappa}} \,.
\end{equation}

Since the inflationary solution corresponds to a saddle fixed point, it is possible to find initial conditions that do not yield to slow-roll dynamics, i.e. inflation does not last enough time. For example, we can regard $x_i \neq z_i$, i.e. we choose a nonzero speed for the field as an initial condition. As expected, the main effect is a drastic reduction in the number of $e$-folds while generating bigger values for $\gamma_i$. For the same value of $\alpha = 2 \times 10^{6}$ and being careful about $y_i^2 > 0$ [remember the constraint in Eq. (\ref{inf constraint}], we consider the following initial values:
\begin{equation}
x_i = 0.070651 \,,\quad z_i = 0.0706 \,,
\end{equation}
yielding to
\begin{equation}
\gamma_i = 0.58811 \,,\quad \epsilon_i = 0.0158458 \,,
\end{equation}
and corresponding to
\begin{equation} \label{GF efolds 2}
N \approx 49 \,,
\end{equation}
showing the great impact that a slight change in the speed of the field ($\dot{\psi}_i \approx 7.2 \times 10^{-5} H_i$) has on the evolution of the system. This feature is important since, at perturbative level, there are bounds on the possible values of $\gamma$ (see Ref. \cite{Maleknejad:2011sq}), and as shown in Ref. \cite{Namba:2013kia}, the scalar perturbations present a strong tachyonic instability for $\gamma < 2$. Nonetheless, we do not care about these cumbersome features since we are only interested in the dynamical behaviour of the system at the background level.

From these numerical results, we also can give a rough estimation of the values of the parameters $\kappa$ and $g$ noticing that
\begin{equation} \label{GF H}
\frac{H}{g} = \frac{\sqrt{2} z^2}{y} \, m_\text{P} \,.
\end{equation}
Although the energy scale of inflation is not known yet, the preferred scale is $H \lesssim 10^{-5} m_\text{P}$. Then, supposing $H_i = 3.5 \times 10^{-5} m_\text{P}$ and using the first set of initial conditions we get
\begin{equation} \label{GF parameters}
g \approx 7.6 \times 10^{-6} \,, \quad \kappa \approx 3.4 \times 10^{16} \,\, m_\text{P}^{-4} \,.
\end{equation}
These parameters were estimated in Ref. \cite{Maleknejad:2011sq} after a linear perturbation treatment of the theory in order to agree with the observational data available at that time \cite{Komatsu:2010fb}. However, here we have shown that it is possible to estimate the order of magnitude of them through a classical dynamical system approach in a much simpler way.

%%%%%%%%%%%%%%%%%%%%%%%%%%%%%%%%%%%%

\subsection{Massive case} \label{Massive GF}

At the background level, the gaugeflation scenario can solve the classical cosmological problems by producing an inflationary phase which lasts enough number of $e$-folds. Nonetheless, the main test of any inflationary model relies in its observational signatures, which are imprinted on the CMB data. In Ref. \cite{Namba:2013kia}, it was shown that some difficulties arise at the linear perturbative level: for $\gamma < 2$, the scalar perturbations show a strong tachyonic instability, and, in the stable region $\gamma > 2$, any set of initial conditions does not exist that preserves the relation between the tensor-to-scalar ratio $r$ and the spectral index $n_s$ in the confidence region allowed by the results of Planck 2013 \cite{Planck:2013jfk}. Then, it is clear that the theory requires some modifications in order to not be discarded by observations. 

In Ref. \cite{Nieto:2016gnp} was proposed ``massive gaugeflation", where an explicit gauge-symmetry breaking mass term was considered. In that work, some of the classical inflationary trajectories of the model were studied, showing that the introduction of the mass term does not harm the good features of the original gaugeflation model while having the potential to modify the $n_s \, \text{vs} \, r$ relation. The linear perturbation theory was carried out in Ref. \cite{Adshead:2017hnc}, where it was shown that the dynamics remain unstable for $\gamma < 2$, nonetheless the model can produce observationally viable spectra in the stable region, according to Planck 2013 \cite{Planck:2013jfk}.

In this case $m \neq 0$, implying $\omega \neq 0$, and so the dynamical system is given by the full set of Eqs. (\ref{inf x eq})-(\ref{inf z eq}). The system has again only one physically acceptable point, where $x$ and $z$ take the same values in Eq. (\ref{inf fixed point}) and $w = 0$. Since $w = 0$, this point has the same properties as in the massless case: it exists for all $\alpha$, the slow-roll parameter is the same in Eq. (\ref{GF slowroll}), and slow-roll solutions require a large $\alpha$. This is not a surprise since the mass term does not contribute to the existence of an accelerated expansion. Note that in this case, we cannot fully specify the initial conditions for inflation only by fixing $\alpha$ since $\omega$ is not given in terms of $\alpha$. 

\begin{figure}[t!]
\centering
\includegraphics[width=\linewidth]{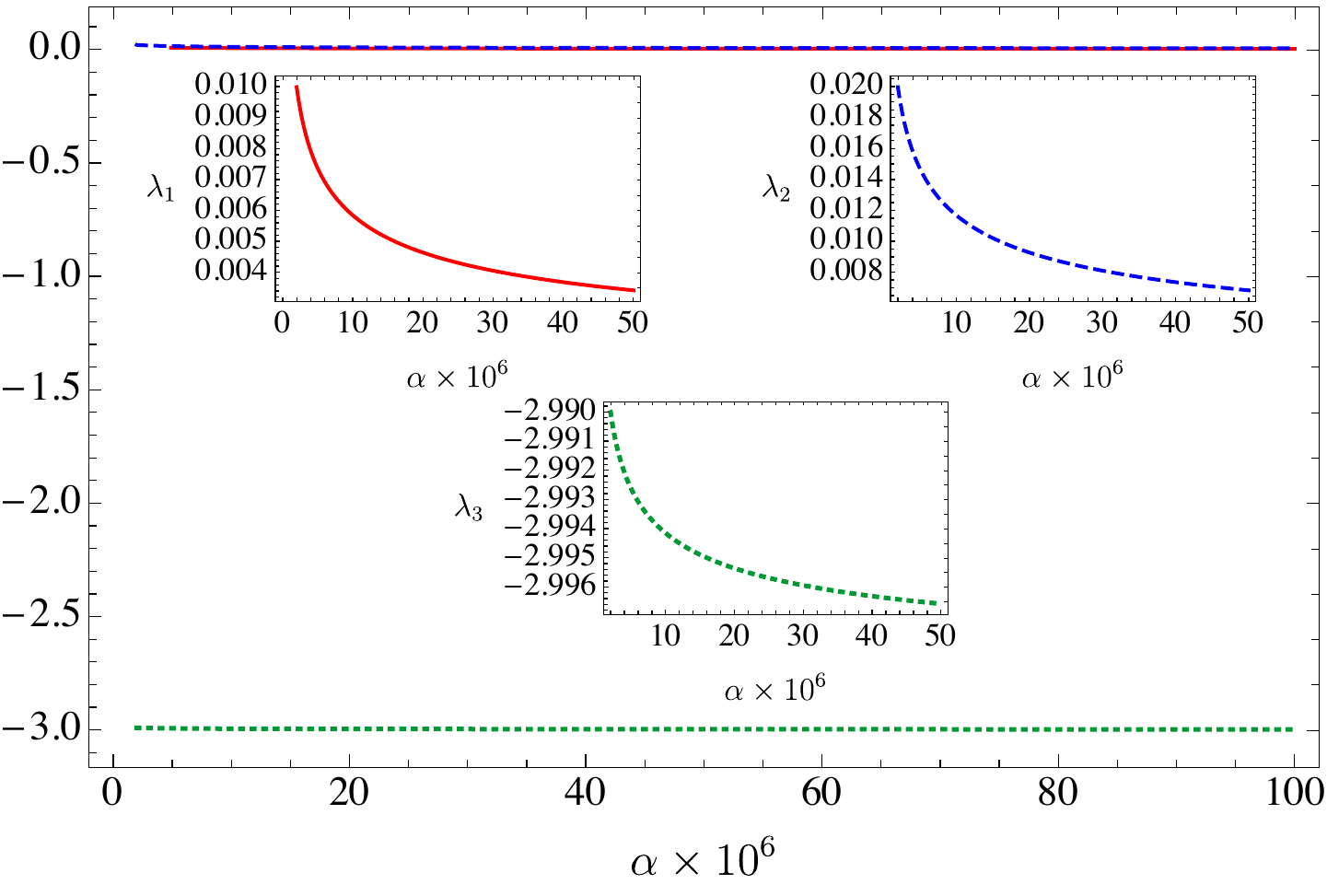}
\caption{Eigenvalues of the Jacobian matrix. Since $\lambda_1 > 0 , \lambda_2 > 0$ and $\lambda_3 < 0 $ in the region where $\epsilon_i < 10^{-2}$, corresponding to inflationary solutions.}
\label{MGF eigenvalues}
\end{figure}

The linear stability analysis shows that this point is hyperbolic as well since there are no vanishing eigenvalues. In Fig. \ref{MGF eigenvalues} we plot these eigenvalues as functions of the parameter $\alpha$ in the region where slow-roll solutions exist. We note that two eigenvalues are positive, $\lambda_1 > 0 \, \text{and} \, \lambda_2 > 0$, while the other one is negative, $\lambda_3 < 0$, therefore the fixed point is a saddle in the phase space $(x, w, z)$, as expected. We verified that $\lambda_1 \rightarrow 0$, $\lambda_2 \rightarrow 0$ and $\lambda_3 \rightarrow -3$ when $\alpha \rightarrow \infty$.

The eigenvalues $\lambda_2$ and $\lambda_3$ in Fig. \ref{MGF eigenvalues} are the same $\lambda_1$ and $\lambda_2$ in Fig. \ref{GF eigenvalues}, respectively. The new eigenvalue is $\lambda_1$. The eigenvector associated with $\lambda_1$ is $(0, 1, 0)$ in the space $(x, w, z)$, meaning that the variable $w$ runs away from 0 to a greater value during inflation. This is expected since at the end of inflation the $\kappa$-term decays to zero and the Yang-Mills term or the mass term becomes dominant.

As mentioned before, the parameter $\alpha$ does not fully specify the initial conditions (unless $m = 0$), then we have to investigate the impact that the mass term, encoded in the function $\omega$ in Eq. (\ref{inf omega}), has on the inflationary phase. 

\begin{figure}[!t]
\centering
\includegraphics[height = 5.2cm, width=0.9\linewidth]{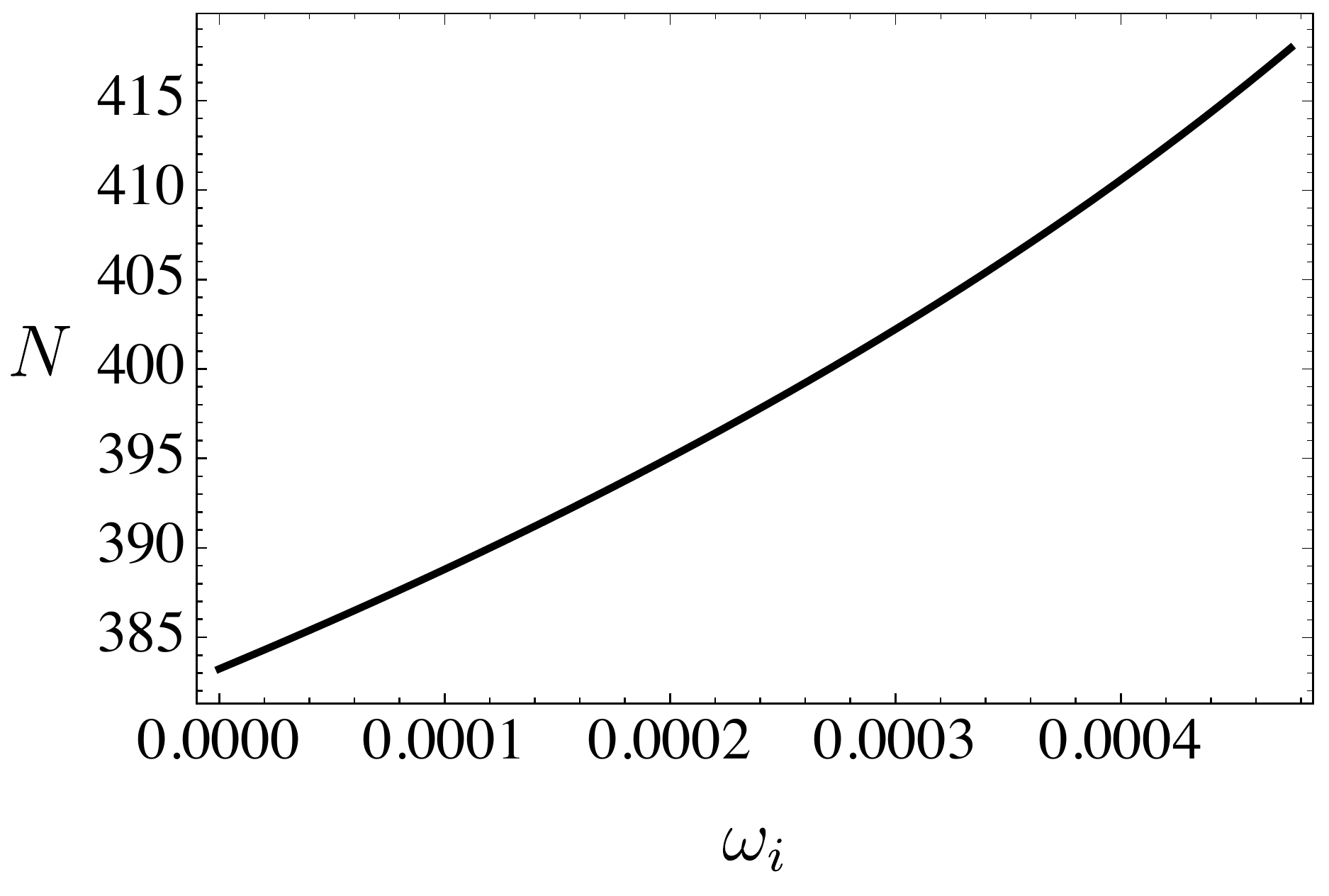}
\caption{Number of $e$-folds in the function of $\omega_i$. The mass term increases the length of the inflationary phase.}
\label{MGF efolds plot}
\end{figure}

Using Eq. (\ref{inf omega}), we can write the constraint in Eq. (\ref{inf constraint}) as
\begin{equation} \label{MGF y 2}
y^2 = 1 - x^2 (1 + 4 \alpha z^4) - \omega^2 z^2 \,.
\end{equation}
From the last equation, it is clear that once $\alpha, x$, and $z$ are fixed, $\omega$ cannot take arbitrary values since it could push the variable $y$ to complex values, which is physically unacceptable. Now, using the first set of initial conditions of the massless case:
\begin{equation}
\alpha = 2 \times 10^{6} \,,\quad x_i = z_i = 0.0706517 \,,
\end{equation}
we get
\begin{equation} \label{MGF omega constraint}
\omega_i < 0.00047539 \,,
\end{equation}
for $y_i^2 > 0$. Since $y_i$ depends on the choice of $\omega_i$, we note that $\gamma_i$ will depend on this function through Eq. (\ref{inf gamma}). Therefore the number of $e$-folds can be given only in terms of $\omega_i$. As it can be seen in Fig. \ref{MGF efolds plot}, the case $\omega_i = 0$ corresponds to $N \approx 382$ agreeing perfectly with the result given in Eq. (\ref{GF 383 efolds}). From this figure, it is clear the effect that the mass term has on the model: it increases the length of the inflationary phase. Furthermore, by using Eq. (\ref{inf omega}), Eq. (\ref{GF H}) and the Friedmann constraint in Eq. (\ref{MGF y 2}) we can write $N$ in terms of the mass $m$ which yields to the same conclusion (see Appendix \ref{N vs m} for a detailed derivation). It is worth mentioning here the results given in Ref. \cite{Nieto:2016gnp}. In that work, the authors claim that the mass ``reduces" the length of inflation, which they show in their Fig. 3. However, this opposite conclusion to ours is because the authors of Ref. \cite{Nieto:2016gnp} plot the number of $e$-folds as a function of $\sqrt{\omega_i}$ (fixing $H_i$) ignoring the relation between this function and $\gamma_i$ through Friedmann equations, i.e. $\gamma_i$ cannot be fixed if $\omega_i$ is varying.

\begin{figure}[t!]
%\subfloat[]
{
\centering
\includegraphics[width=\columnwidth]{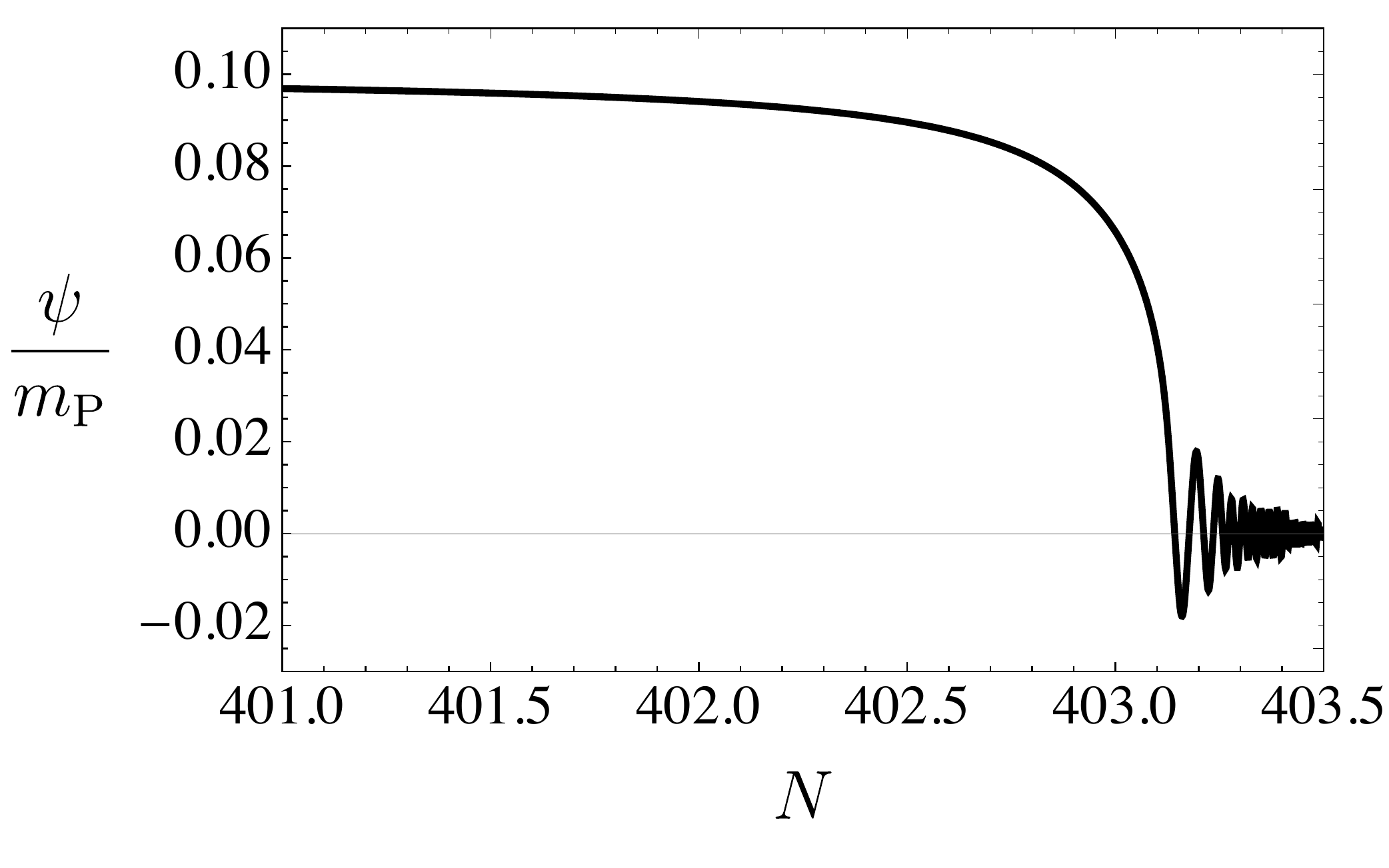}
%\label{MGF field}
}
\vspace{0.05cm}
%\subfloat[]
{
  \centering
  \includegraphics[width=0.92\columnwidth]{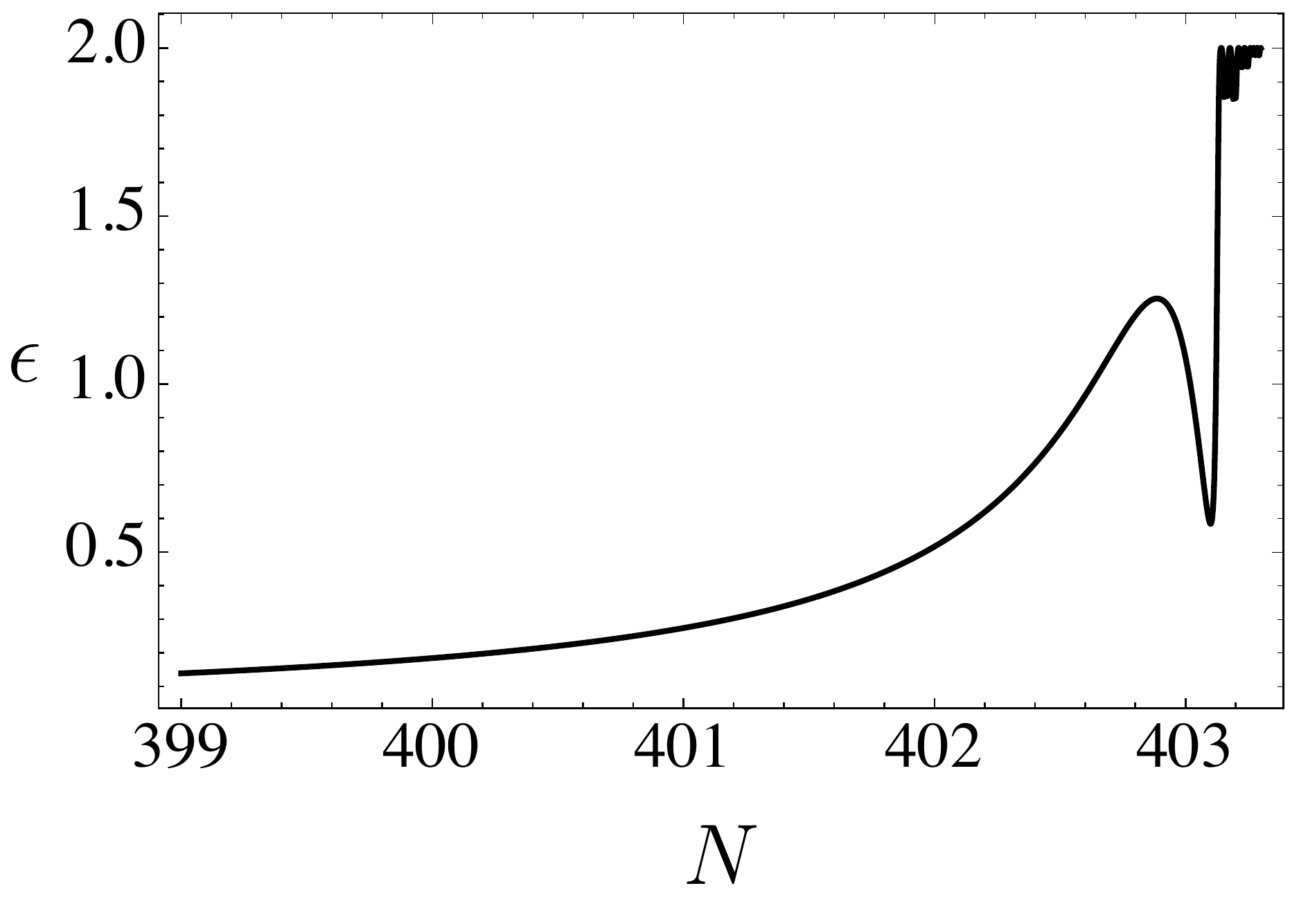}
%  \label{MGF epsilon}
}
\caption{Evolution of the gauge field (upper) and the slow-roll parameter (lower). The field $\psi = \phi / a$ and the slow-roll parameter $\epsilon$ are nearly constant during inflation. The field suddenly decays while $\epsilon$ reaches its upper bound around $N = 403$ as stated in Eq. (\ref{MGF efolds}). The parameter $\epsilon$ presents another peak around $1$ due to the presence of the mass term.}
\label{inf massive}
\end{figure}

As a complement to our qualitative and analytical results, we numerically integrate the full dynamical system in Eqs. (\ref{inf x eq})-(\ref{inf z eq}). Choosing 
\begin{equation}
\omega_i = 3 \times 10^{-4} \,,
\end{equation}
we get
\begin{equation}
\gamma_i \approx 1.8 \times 10^{-4} \,,\quad \epsilon_i = 9.9 \times 10^{-3} \,,
\end{equation}
and the expected number of $e$-folds
\begin{equation} \label{MGF efolds}
N \approx 402 \,.
\end{equation}

As can be seen in the upper part of Fig. \ref{inf massive}, for the chosen initial conditions, the gauge field is nearly constant during inflation, then it suddenly falls off around $N = 403$ and starts to oscillate, showing perfect matching with the analytical results. In the lower part of Fig. \ref{inf massive}, the slow-roll parameter grows from a nearly zero value until $\epsilon = 1$ around $N = 402$ where slow-roll inflation ends. At this stage, the Universe is dominated by the mass term but soon the Yang-Mills term becomes dominant and the slow-roll parameter oscillates below its upper bound $\epsilon = 2$.

It is also possible to consider greater values for $\omega_i$, by regarding $x_i \neq z_i$. As in the massless gaugeflation model, this implies a nonzero speed for the field and it has the same effect here: it reduces the length of inflation. However, this does not cancel the effect of the mass term increasing this length as shown in the following. Taking the second set of initial conditions used in the massless case,
\begin{equation}
\alpha = 2 \times 10^{6} \,,\quad x_i = 0.070651 \,,\quad z_i = 0.0706 \,,
\end{equation}
we get 
\begin{equation}
\omega_i < 0.584146 \,,
\end{equation}
for $y_i^2 > 0$ from Eq. (\ref{MGF y 2}). So, choosing $\omega_i = 0.5$ we get $N \approx 69$  as the expected length of inflation although $\epsilon_i \approx 1.3 \times 10^{-2}$. Remember that with these initial conditions, the massless model predicts $N \approx 49$ ruling it out as a viable inflationary solution.

From this analysis we can also estimate the values of the parameters $\kappa$ and $g$, through Eq. (\ref{GF H}). Using the energy scale $H_i = 3.5 \times 10^{-5} m_\text{P}$ and using the second set of initial conditions we get
\begin{equation}
g \approx 10^{-4} \,,\quad \kappa \approx 1.9 \times 10^{14} \,\, m_\text{P}^{-4} \,,
\end{equation}
which are roughly of the order to the values estimated in Eq. (\ref{GF parameters}) for the massless model.

%%%%%%%%%%%%%%%%%%%%%%%%%%%%%%%%%%%%

\subsubsection*{Possible consequences at the perturbative level}

From the analysis above we learn that once the constant $\alpha$, the magnitude of the field $\psi$, and its speed $\psi' = \dot{\psi}/H$ are fixed, the parameters $\gamma$ and $\omega$ are related by the Friedmann constraint in Eq. (\ref{inf constraint}), in such a way that one cannot fix one of them while varying the other one. To elucidate the possible impact of the relation between $\gamma$ and $\omega$, we take a minimal example coming  from the results obtained in Ref. \cite{Adshead:2017hnc} where the linear perturbation of the massive gaugeflation model was worked out. Assuming $\alpha = 10^9,\, x = 0.02509638,\, z = 0.02508$ and taking into account that $y$ has to be real and $\gamma > 2$ in order to avoid tachyonic instabilities, we get the bound
\begin{equation}
\omega < 2.14366 \,.
\end{equation} 
The late time decay rate of the scalar power spectrum is given by [see Eq. (3.58) of Ref. \cite{Adshead:2017hnc}]

\begin{equation}\label{eq:etascal}
n_{\text{scal}} = \epsilon \left[ 1 + \frac{M^2-6}{2\gamma + M^2 + 2}\right] \, .
\end{equation} 
 
\begin{figure}[t!]
{
\centering
\includegraphics[width=0.98\columnwidth]{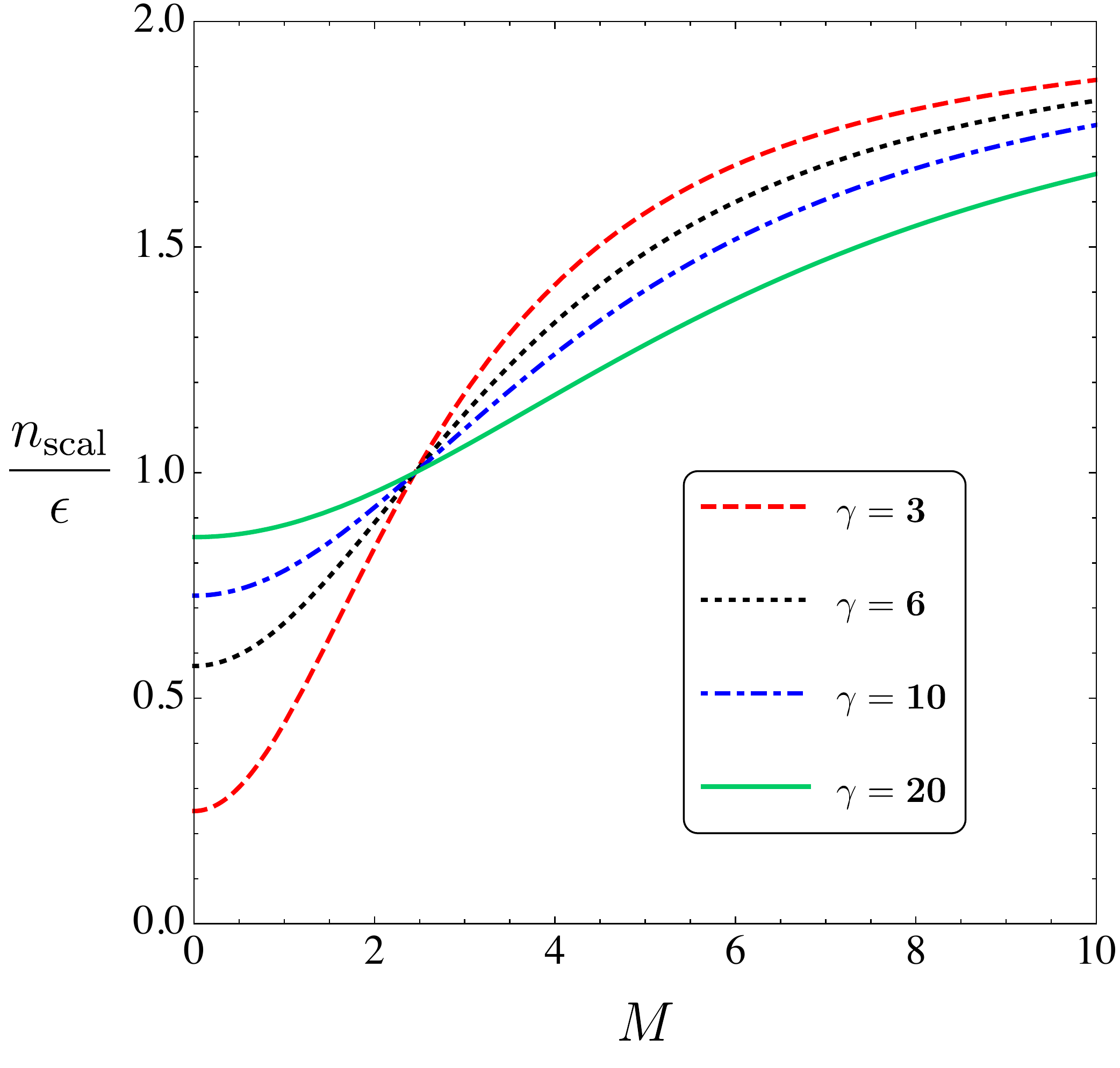}
}
\vspace{0.05cm}

{
  \centering
  \includegraphics[width=\columnwidth]{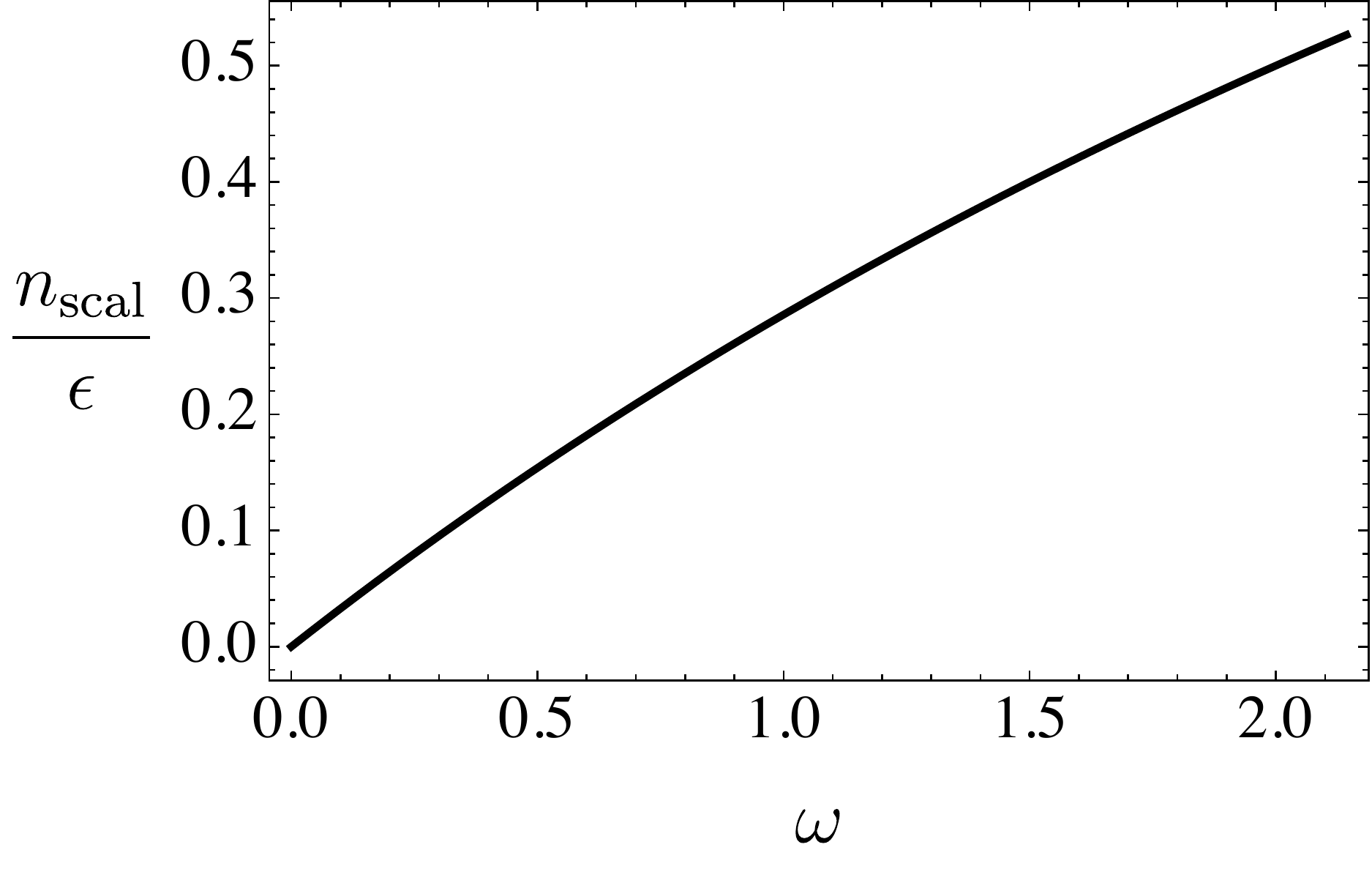}
}
\caption{Plot of the late-time decay of the scalar power spectrum $n_{\text{scal}}/ \epsilon$ as a function of the mass parameters given by Eq. (3.58) of \cite{Adshead:2017hnc}. Upper: result obtained when $\gamma$ is fixed while $M = m / H$ can vary (see Fig. 5 of Ref. \cite{Adshead:2017hnc}). Lower: result obtained when the relation between $\gamma$ and $\omega = M^2$ is considered.}
\label{nscal}
\end{figure}

In Fig. \ref{nscal} we plot $n_{\text{scal}}$ normalized to $\epsilon$ with respect to the mass parameter. The upper part of Fig. \ref{nscal} is the plot shown in Ref. \cite{Adshead:2017hnc}, where $\gamma$ is fixed and $M = m / H$ varies. The lower part of Fig. \ref{nscal} shows the result obtained when taking into account the relation between $\gamma$ and $\omega$, which of course corresponds to only one curve. We also found some differences between the plots of the chirality parameter [Eq. (3.114) of Ref. \cite{Adshead:2017hnc}] and those made by us, considering the relation between $\gamma$ and $\omega$. We want to stress that the perturbative analysis of the model is out of the scope of this work, and although the dependence between the parameters $\gamma$ and $\omega$ has the potential to modify the scalar spectral index or the tensor-to-scalar ratio (as we see from the expression for $n_{\text{scal}}$), a full treatment of perturbations for the massive case is needed. We leave this complete examination for a future work, where we also plan to compare with the results given in  Ref. \cite{Adshead:2017hnc}, and also to see if the massive gaugeflation proposal can be (or not be) in agreement with the observational bounds.

%%%%%%%%%%%%%%%%%%%%%%%%%%%%%%%%%%%%

\section{Dark energy from gauge fields} \label{dark energy}

In Ref. \cite{Mehrabi:2017xga} the massless case of the action in Eq. (\ref{gauge action}) was studied in the context of dark energy. There it was shown that, for several sets of initial conditions and parameters, the dynamics of the non-Abelian gauge field can account for the late-time accelerated expansion of the Universe. Here we extend the model by considering the effect of the mass term in the cosmological evolution and use a dynamical system approach to fully describe the late-time behaviour of the expansion. In particular, we explicitly show that dark energy domination is indeed an attractor and give the full parameter space of the theory, complementing the results given in Ref. \cite{Mehrabi:2017xga}.

In order to reproduce the known expansion history of the Universe, let us modify the density and pressure in Eq. (\ref{rho p}) as
\begin{equation}
\rho = \rho_{\text{DE}} + \rho_m + \rho_r \,,\,  
\end{equation}
\begin{equation}
p = p_{\text{DE}} + \frac{1}{3} \rho_r \,, 
\end{equation}
where $\rho_m$ is the density of dust ($p_m = 0$), $\rho_r$ is the density of radiation ($p_r = \rho_r / 3$), and $\rho_{\text{DE}}$ and $p_\text{DE}$ are given by Eq. (\ref{rho p}). The Friedmann equations in Eqs. (\ref{inf H2}) and (\ref{inf Hdot}) become
\begin{multline} \label{de H2} 
3 m_\text{P}^2 H^2 = \frac{3}{2} \left[ \frac{\dot{\phi}^2}{a^2} +  \frac{g^2 \phi^4}{a^4} +  \kappa g^2 \frac{\phi^4 \dot{\phi}^2}{a^6}  + \frac{m^2 \phi^2}{a^2} \right] \\
 + \rho_m + \rho_r \,, 
 \end{multline}
 \begin{multline}\label{de Hdot}
2 m_\text{P}^2 \dot{H} = - 2 \left[ \frac{\dot{\phi}^2}{a^2} + \frac{g^2 \phi^4}{a^4} + \frac{1}{2} \frac{m^2 \phi^2}{a^2}  \right. \\ 
\left. + \frac{1}{2} \rho_m + \frac{2}{3} \rho_r \right] \,. 
\end{multline}
Since the barotropic fluids do not introduce new terms to the gauge field equation of motion, it is given again by  Eq. (\ref{field eom}). These equations are complemented by the usual continuity equations for the barotropic fluids
\begin{equation} \label{de cont eqs}
\dot{\rho}_m + 3H \rho_m = 0 \, , \quad \dot{\rho}_r + 4H \rho_r = 0 \, .
\end{equation}

Proceeding as in the inflationary case, from the Friedmann equation in Eq. (\ref{de H2}) we define the dimensionless variables defined in Eq. (\ref{inf variables}) and the density parameters for radiation and matter as
\begin{equation} \label{de variables}
\Omega_r \equiv \frac{\rho_r}{3\,m_\text{P}^2 H^2}\, , \quad \Omega_m \equiv \frac{\rho_m}{3\,m_\text{P}^2 H^2} \,.
\end{equation}
Hence Eq. (\ref{de H2}) becomes a constraint from which we can write $\Omega_m$ in terms of the other variables as
\begin{equation} \label{de constraint}
\Omega_m = 1 - x^2 (1 + 4 \alpha z^4) - y^2 - w^2 - \Omega_r \,.
\end{equation}
Using the equation of motion in Eq. (\ref{field eom}), the continuity equation in Eq. (\ref{de cont eqs}) for radiation fluid, and the constraint in Eq. (\ref{de constraint}), the autonomous set of equations reads
\begin{align}
x' &= x q - \left(1 + 4 \alpha z^4\right)^{-1} \Big[ 2 \frac{y^2}{z} + 8 \alpha x^2 z^3 \nonumber \\
&+ x \left(1 - 12 \alpha z^4 \right) + \frac{w^2}{z} \Big], \label{de  x} \\
y' &= y \left( 2 \frac{x}{z} + q -1 \right) \,, \\
w' &= w \left( \frac{x}{z} + q \right) \, \\
z' &= x - z \,, \\
\Omega'_r &= 2 \Omega_r \left(q - 1\right) \,, \label{de r}  
\end{align}
where the deceleration parameter $q \equiv -1 - \dot{H}/H^2$ is obtained from Eq. (\ref{de Hdot}) as
\begin{equation} \label{DE deceleration}
q = \frac{1}{2} \left[1 + x^2 \left(1 - 12 \alpha z^4 \right) + y^2 - w^2 + \Omega_r \right] \, .
\end{equation}
From this deceleration parameter, we can define the effective equation of state $w_\text{eff} \equiv \frac{2q - 1}{3} $ which in terms of the dimensionless variables reads 
\begin{equation} \label{eq:weff}
w_\text{eff} = \frac{1}{3} \left[ x^2 \left(1 - 12 \alpha z^4 \right) + y^2 - w^2 + \Omega_r \right] \,,
\end{equation}
which completely characterizes the evolution of the average scale factor $a(t)$. The dark sector is characterized by the density parameter \mbox{$\Omega_{\text{DE}} \equiv \rho_{\text{DE}} / \left( 3 m_{\text{P}}^2 H^2 \right)$} and the equation of state \mbox{$w_{\text{DE}} \equiv p_{\text{DE}} / \rho_{\text{DE}}$}:
\begin{equation}
\Omega_{\text{DE}} = x^2 \left( 1 + 4 \alpha z^4 \right) + y^2 + w^2 \,,
\end{equation}
\begin{equation}
w_{\text{DE}} = \frac{1}{3} \frac{x^2 \left(1 - 12 \alpha z^4 \right) + y^2 - w^2}{x^2 \left(1 + 4 \alpha z^4 \right) + y^2 + w^2} \,.
\end{equation}

%%%%%%%%%%%%%%%%%%%%%%%%%%%%%%%%%%%%

\subsection{Fixed points}

In what follows, we study the fixed points relevant for the cosmological evolution, namely, the radiation era ($\Omega_r \simeq 1,  w_\text{eff}  \simeq 1/3$), the matter era ($\Omega_m \simeq 1, w_\text{eff} \simeq 0$), and the dark energy era ($\Omega_{\text{DE}} \simeq 1, w_\text{eff} < -1/3$). 

\begin{itemize}
\item (\emph{R}) Radiation dominance 
\end{itemize}
\begin{equation}
x = 0 \,,\, y = 0 \,,\, w = 0 \,,\, z = 0 \,,\, \Omega_m = 0 \,,
\end{equation}
with $\Omega_{\text{DE}} = 0$, $w_{\text{DE}}$ undetermined and $\Omega_r = 1$.

\begin{itemize}
\item (\emph{M}) Matter dominance 
\end{itemize}
\begin{equation}
x = 0 \,,\, y = 0 \,,\, w = 0 \,,\, z = 0 \,,\, \Omega_r = 0 \,,
\end{equation}
with $\Omega_{\text{DE}} = 0$, $w_{\text{DE}}$ undetermined and $\Omega_m = 1$.

\begin{itemize}
\item (\emph{S}) Scaling matter-dark energy
\end{itemize}
\begin{equation}
x = z = \frac{1}{\sqrt{2 \,} \sqrt[4]{3 \alpha}} \,,\, y = 0 \,,\, w = 0 \,,\, \Omega_r = 0 \,,
\end{equation}
with $\Omega_{\text{DE}} = 2 / \left( 3 \sqrt{3 \alpha} \, \right)$, $w_{\text{DE}} = 0$  and $\Omega_m = 1 - \Omega_{\text{DE}}$.

In order to have \mbox{$\Omega_m > 0 $} we need to consider \mbox{$ \alpha > 0.148148 \,. $} This point corresponds to an effective matter epoch since $w_{\text{eff}} \simeq 0$. From the CMB constraint given by Planck \cite{Ade:2015rim}, the density parameter of dark energy is constrained to be $\Omega_{\text{DE}} < 0.02$ around the redshift $z_r=50$,\footnote{We will denote the redshift by $z_r$ to avoid confusion with the variable $z$ in the dynamical systems.} which implies \mbox{$ \alpha > 370.37 \,.$} For a proper matter epoch, $\Omega_m$ has to be the dominant component in the energy budget. For instance, requiring $ 0.999 < \Omega \leq 1 \,,$ we find \mbox{$ \alpha > 1.48148 \times 10^5 \,.$} Aside of the exact value of $\alpha$, we can conclude that $\alpha$ has to be large.

\begin{itemize}
\item (\emph{DE}) Dark energy dominance
\end{itemize}
\begin{equation} \label{DE attractor}
x = z = \sqrt{\frac{\sqrt[3]{\alpha} f_{\alpha}^2 - \sqrt[3]{3}}{2 f_{\alpha} \sqrt[3]{9\alpha^2} }} \,,\, y = 0 \,,\, w = 0 \,,\, \Omega_r = 0 \,,
\end{equation}
with $\Omega_{\text{DE}} = 1$, $\Omega_m = 0$ and
\begin{equation} \label{de eos}
w_{\text{DE}} = - 1 + \frac{2 f_\alpha}{9} \left[ \sqrt[3]{\frac{3}{\alpha}} + \sqrt[3]{\frac{\alpha}{3}} f_\alpha \left(18 -  f_\alpha^3  \right) \right] \,.
\end{equation}
This point corresponds to an accelerated expansion solution if $ w_{\text{DE}} < -1/3 \,,$ which implies that $\alpha > 1$. Since observations favor $w_{\text{DE}} \approx -1$ \cite{Ade:2015xua}, from Eq. (\ref{de eos}), for example, $ -1 \leq w_{\text{DE}} < - 0.99 \,$ implies $ \alpha > 5.88148 \times 10^5 \,. $ We conclude that the dark energy component can behave in agreement with the observational bounds for large $\alpha$. 

%%%%%%%%%%%%%%%%%%%%%%%%%%%%%%%%%%%%

\subsection{Stability analysis}

In the present case, the fixed points (\emph{R}) and (\emph{M}) require $z = 0$, thence the eigenvalues evaluated in these points could yield to singularities. However, only the sign of the real part of the eigenvalues carries the information about the stability of the point. Therefore, we consider the eigenvalues $\pm \infty$ as a positive or negative eigenvalue. Now, we proceed to discuss the stability of each fixed point by analyzing the sign of the eigenvalues $\lambda_{1, 2, 3, 4, 5}$.

\begin{itemize}
\item (\emph{R}) Radiation dominance
\end{itemize}
\begin{equation}
2, \quad 1, \quad, - \infty, \quad +\infty, \quad + \infty \,.
\end{equation}
This point is a saddle since four eigenvalues are positive and one is negative. The eigenvector associated with $\lambda_2 = 1$ is $(0, 0, 0, 0, 1)$ in the space $(x, y, w, z, \Omega_r)$. Therefore, the point is unstable in the $\Omega_r$ direction, meaning that the variable $\Omega_r$ runs away from its value in the fixed point. 

\begin{itemize}
\item (\emph{M}) Matter dominance
\end{itemize}
\begin{equation}
\frac{3}{2}, \quad -1, \quad, - \infty, \quad +\infty, \quad +\infty \,.
\end{equation}
This point is a saddle since three eigenvalues are positive and two are negative. The eigenvector associated with $\lambda_2 = -1$ is $(0, 0, 0, 0, 1)$, meaning that the radiation contribution is decreasing since $\Omega_r = 0$ is an attractor in the \mbox{$\Omega_r$ direction}. 

\begin{itemize}
\item (\emph{S}) Scaling matter-dark energy
\end{itemize}

\begin{equation}
- \frac{3}{4} \pm \sqrt{\frac{33}{16} - \frac{1}{\sqrt{3\alpha}}} \,,\, - \frac{1}{2} \,,\, \frac{3}{2} \,,\, \frac{3}{2} \,.
\end{equation}
This point is a saddle independently of the value of $\alpha$. The eigenvectors associated with $\lambda_{4, 5} = 3/2$ are $(0, 0, 1, 0, 0)$ and $(0, 0, 0, 1, 0)$, respectively. This means that the point is a repeller in the $y$ and $w$ directions. The numerical solution given below shows that this point is indeed irrelevant in the cosmological dynamics since $w_\text{DE}$ never spends time around this point. 

\begin{itemize}
\item (\emph{DE}) Dark energy dominance
\end{itemize}

\begin{figure}[t!]
\centering
\includegraphics[width=\linewidth]{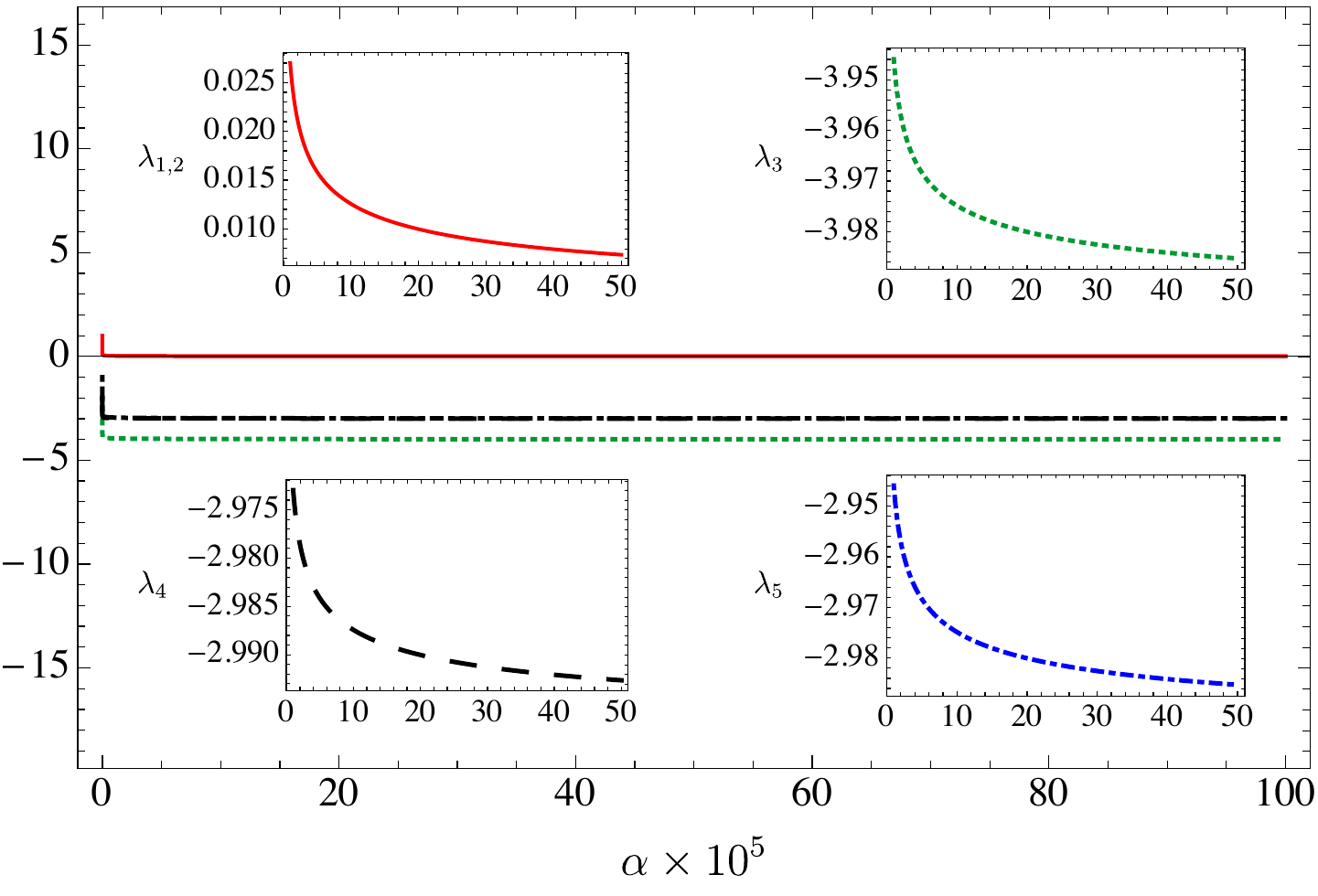}
\caption{The five eigenvalues of the Jacobian matrix. The fixed point is a saddle $\lambda_{1, 2} > 0 \, \text{and} \, \lambda_{3, 4, 5} < 0$. However, this point is an attractor in the relevant physical space. The scale for $\alpha$ is the same for all the plots.}
\label{MGS eigenvalues}
\end{figure}

In this case, the eigenvalues are very long quantities, therefore, we investigate the stability of the point by plotting them  as functions of $\alpha$ for large values of this parameter. In Fig. \ref{MGS eigenvalues}, we can see that $\lambda_{1, 2} > 0$ while the other eigenvalues are negative such that this point corresponds to a saddle in the phase space $(x, y, w, z, \Omega_r)$. We verified that $\lambda_1 \rightarrow 0$, $\lambda_2 \rightarrow 0$, $\lambda_3 \rightarrow -4$, $\lambda_4 \rightarrow -3$ and $\lambda_5 \rightarrow -3$ when $\alpha \rightarrow \infty$. Although in general this point is not an attractor, as usual in dark energy models, this is not a problem for the theory. Moreover, we argue that (\emph{DE}) is indeed an attractor in the relevant physical space. The eigenvector associated with $\lambda_1$ is $(0, 1, 0, 0, 0)$ while the eigenvector of $\lambda_2$ is $(0, 0, 1, 0, 0)$, such that this point is a repeller in the $y$ and $w$ directions, i.e. the variables $y$ and $w$ run away from zero. In this point we have $x = z$, which implies $y \propto \psi^2 / H \propto 1 / H$, given that  $\dot{\psi} = 0$ and $\psi = \text{const}$. Now, since $H$ decreases with the expansion, the variable $y$ increases in the same proportion. Therefore, we realize that $\lambda_1 > 0$ is needed for the theory to be consistent. The same argument follows for $\lambda_2 > 0$ since $w \propto \psi / H$. Therefore, we conclude that this point is a physical attractor but a saddle in the  state space spanned by the chosen variables. This is not surprising since the phase space $(x, y, w, z, \Omega_r)$ is not compact, hence there will not be necessarily a source point and an attractor point in the phase space \cite{Coley:2003mj}. These arguments are further supported by numerical results in the next subsection.

%%%%%%%%%%%%%%%%%%%%%%%%%%%%%%%%%%%%

\subsection{Numerical analysis}

We proceed to solve the autonomous set of equations  (\ref{de x}) to (\ref{de r}) through a numerical integration. Based in the dynamical system analysis, the initial conditions are chosen in the deep radiation era. Explicitly we choose\footnote{Here, the subscript $i$ means that the corresponding quantity is evaluated some time in the deep radiation epoch.}
\begin{equation*}
\alpha = 10^9 \,,\quad x_i = 3 \times 10^{-30} \,,\quad z_i = 1.1 \times 10^{5} \,,
\end{equation*}
\begin{equation}
y_i = 0.001 \,,\quad \Omega_{r_i} = 0.99998 \,.
\end{equation}
The value of $\omega_i$ is constrained by Eq. (\ref{de constraint}). For $\Omega_{m_i} > 0$ we have $\omega_i \equiv m^2/H_i^2 < 1.57025 \times 10^{-15} $, so we choose 
\begin{equation} \label{de omega}
\omega_i = 5 \times 10^{-29} \,,
\end{equation}
corresponding to $\Omega_{m_i} = 1.9 \times 10^{-5}$ at the redshift\footnote{The relation between the number of $e$-folds and the redshift is given by $N = - \text{ln} (1 + z_r)$.} \mbox{$z_r = 2.18 \times 10^{8}$.}

\begin{figure}[t!]
\centering
\includegraphics[width=\linewidth]{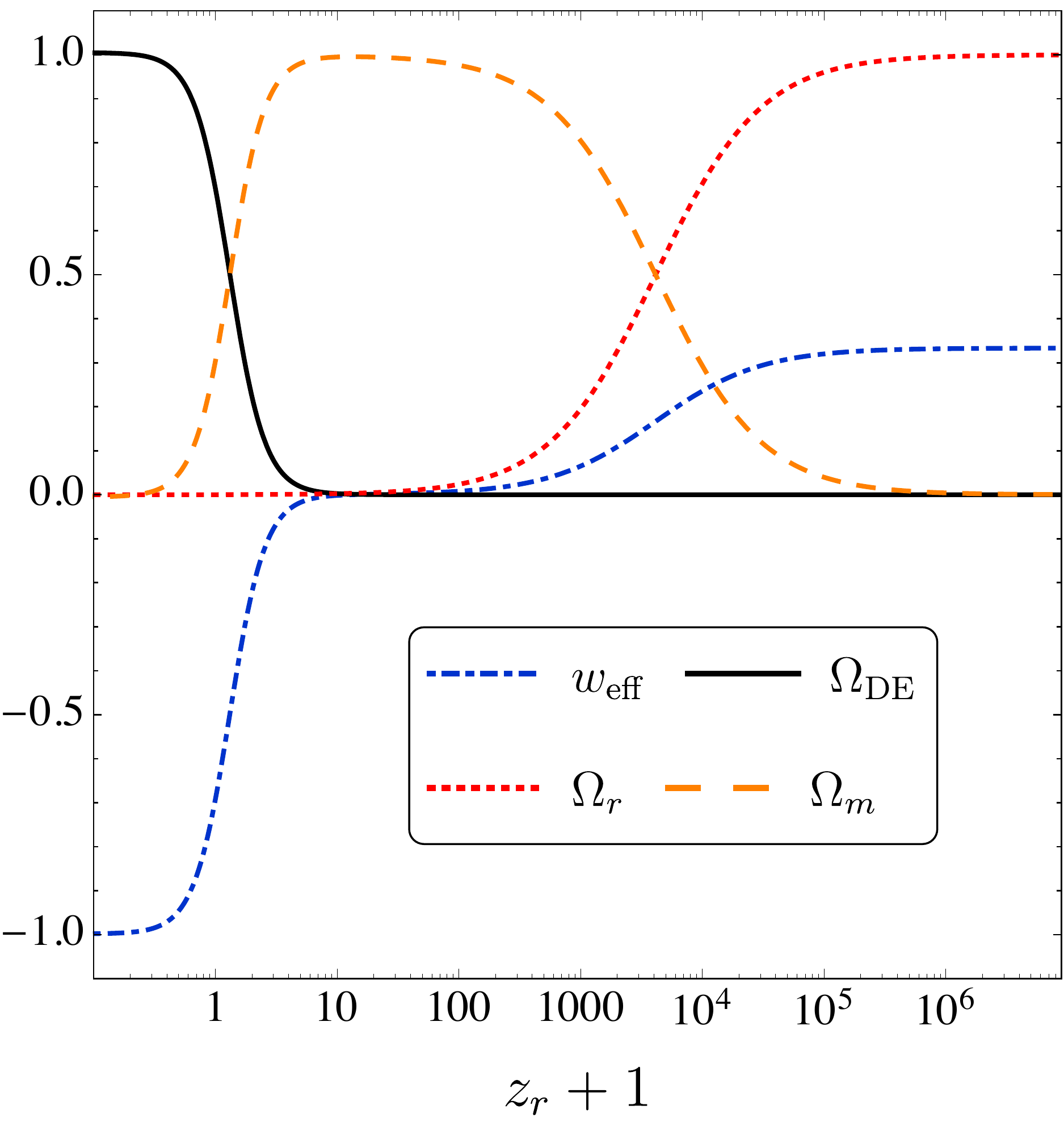}
\caption{Evolution of the density parameters and the effective equation of state during the whole expansion history. The initial conditions were chosen in the deep radiation era at the redshift $z_r = 2.18 \times 10^8$. The Universe passes through radiation dominance at early times (red dotted line), followed by a matter dominance (light brown dashed line), and ends in the dark energy dominance (black solid line) characterized by $w_{\text{eff}} \approx -1$ (blue dot-dashed line).}
\label{de Abundances}
\end{figure}

Using these initial conditions, in Fig. \ref{de Abundances} we plot the density parameters and the effective equation of state as functions of redshift $z_r$. We can see that at early times ($z_r > 10^4$), the dominant cosmic fluid corresponds to radiation (red dotted line), with $w_{\text{eff}} = 1/3$ (blue dot-dashed line). Then, around $z_r = 3200$, we have radiation-matter equality, i.e. $\Omega_r \simeq \Omega_m$. The matter epoch (light brown dashed line) runs from this point to $z_r \approx 0.3$, where $w_{\text{eff}} = 0$. At this point the dark energy component (black solid line) represents the half of the energy budget in the Universe. After this time, the dark energy era begins together with the accelerated expansion characterized by $w_{\text{eff}} \approx - 1$. 

Let us analyze in more detail each of the relevant periods discussed above.

%%%%%%%%%%%%%%%%%%%%%%%%%%%%%%%%%%%%

\subsubsection{Radiation dominated period}

\begin{figure}[t!]
{
\centering
\includegraphics[width=0.98\columnwidth]{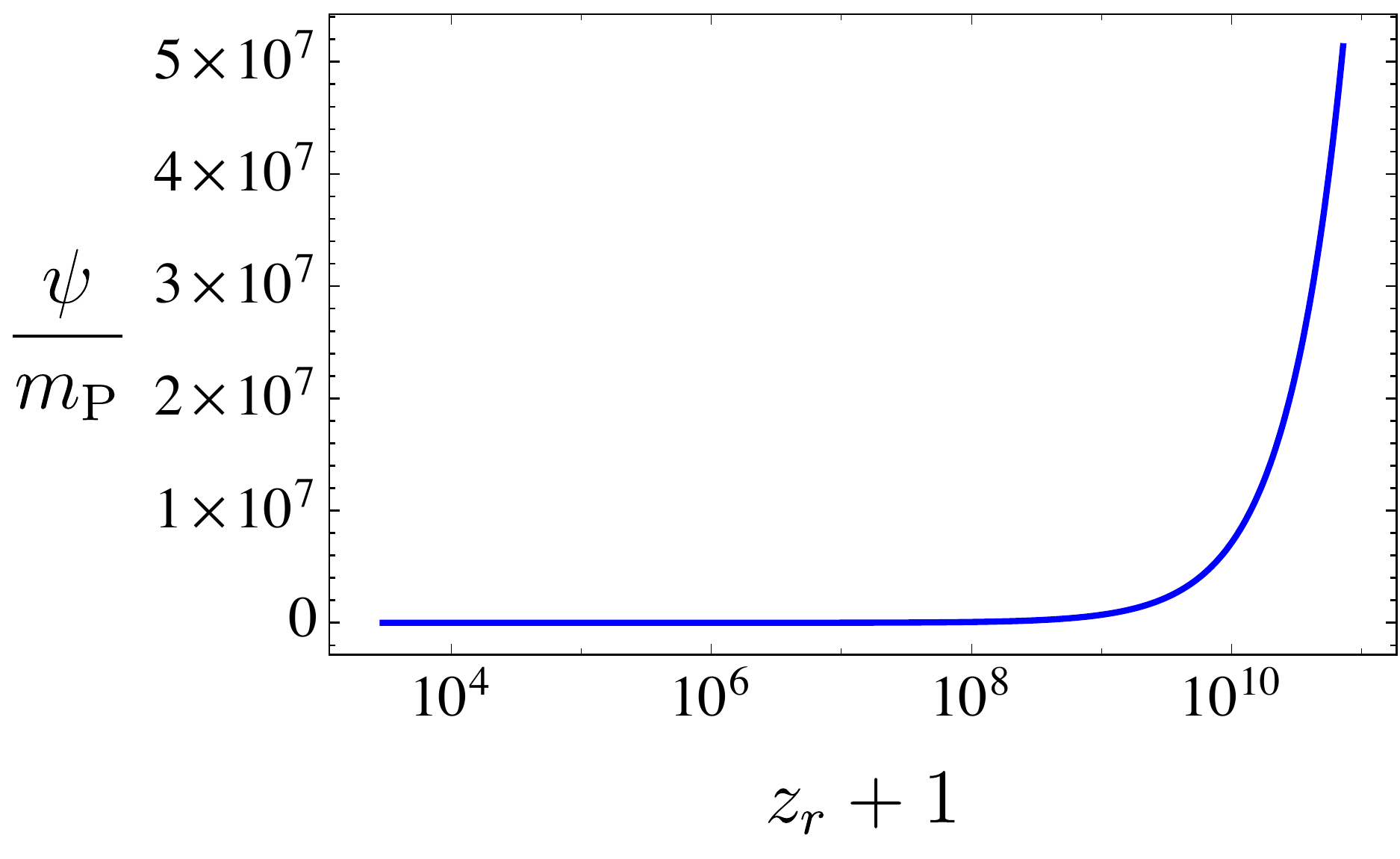}
}
\vspace{0.05cm}

{
  \centering
  \includegraphics[width=0.98\columnwidth]{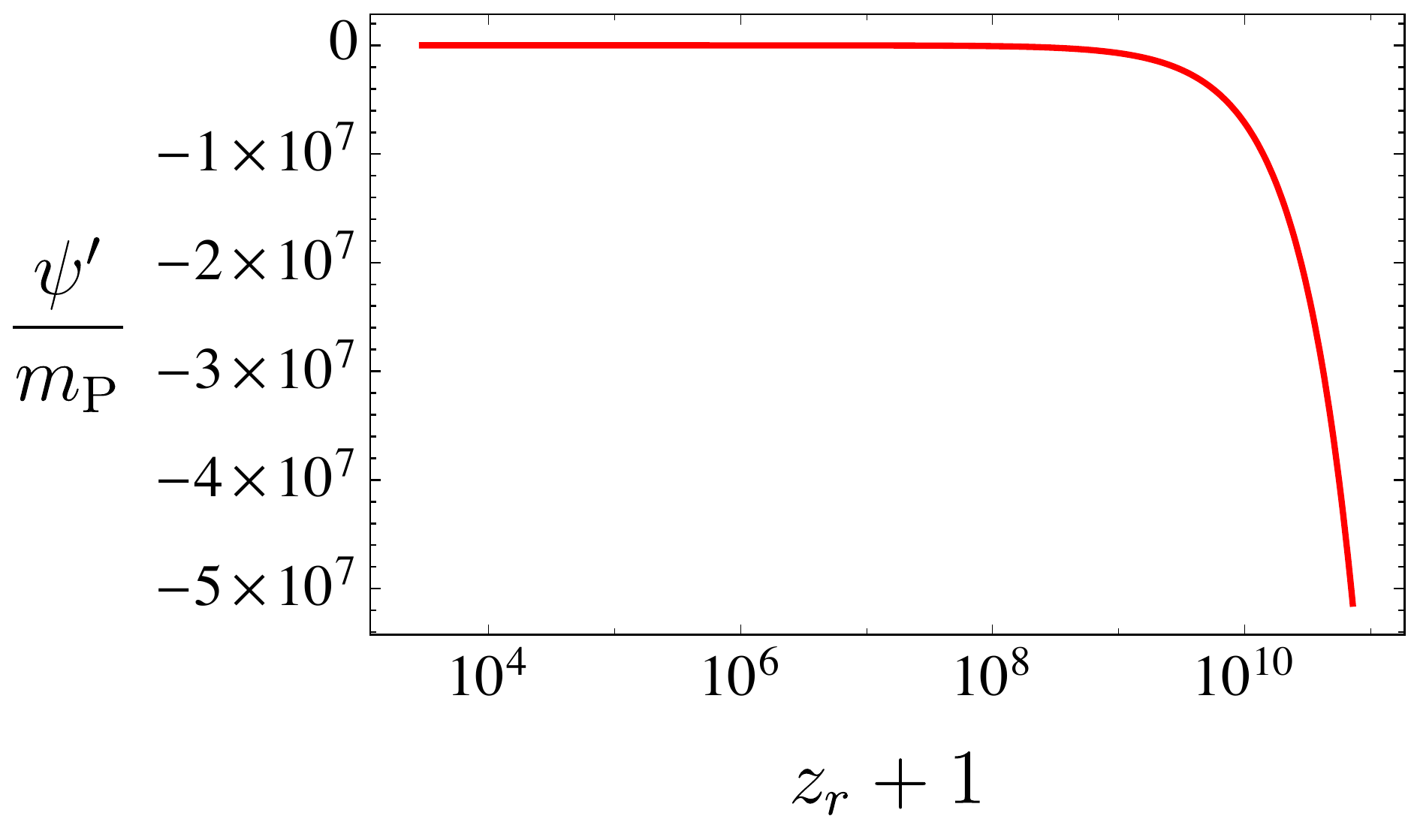}
}
\caption{Evolution of the gauge field (upper), and its speed (lower), during the radiation dominated period. The gauge field decays from a large value in a decelerated way (in magnitude).}
\label{Radiation}
\end{figure}

This period runs from $z_r \simeq 4000$ and on to the past. In Fig. \ref{Radiation} we can see that at early times, the gauge field decays from a large value in a decelerated way (in magnitude). The contribution of early dark energy to this period is $\Omega_{\text{DE}} \approx 4 \times10^{-7}$ well below the big-bang nucleosynthesis (BBN) constraint $\Omega_{\text{DE}} < 0.045$ at $z_r = 3200$ \cite{Bean:2001wt}. During this period, the main contribution to the dark sector comes from the Yang-Mills term. We plot from $z_r \approx 7 \times 10^{10}$ to $z_r = 3000$, such that the total length of the radiation dominated period from the end of inflation to the time of radiation-matter equality is consistent with the constraint given in Ref. \cite{Alvarez:2019ues}.

%%%%%%%%%%%%%%%%%%%%%%%%%%%%%%%%%%%%

\subsubsection{Matter dominated period}

\begin{figure}[t!]
{
\centering
\includegraphics[width=0.98\columnwidth]{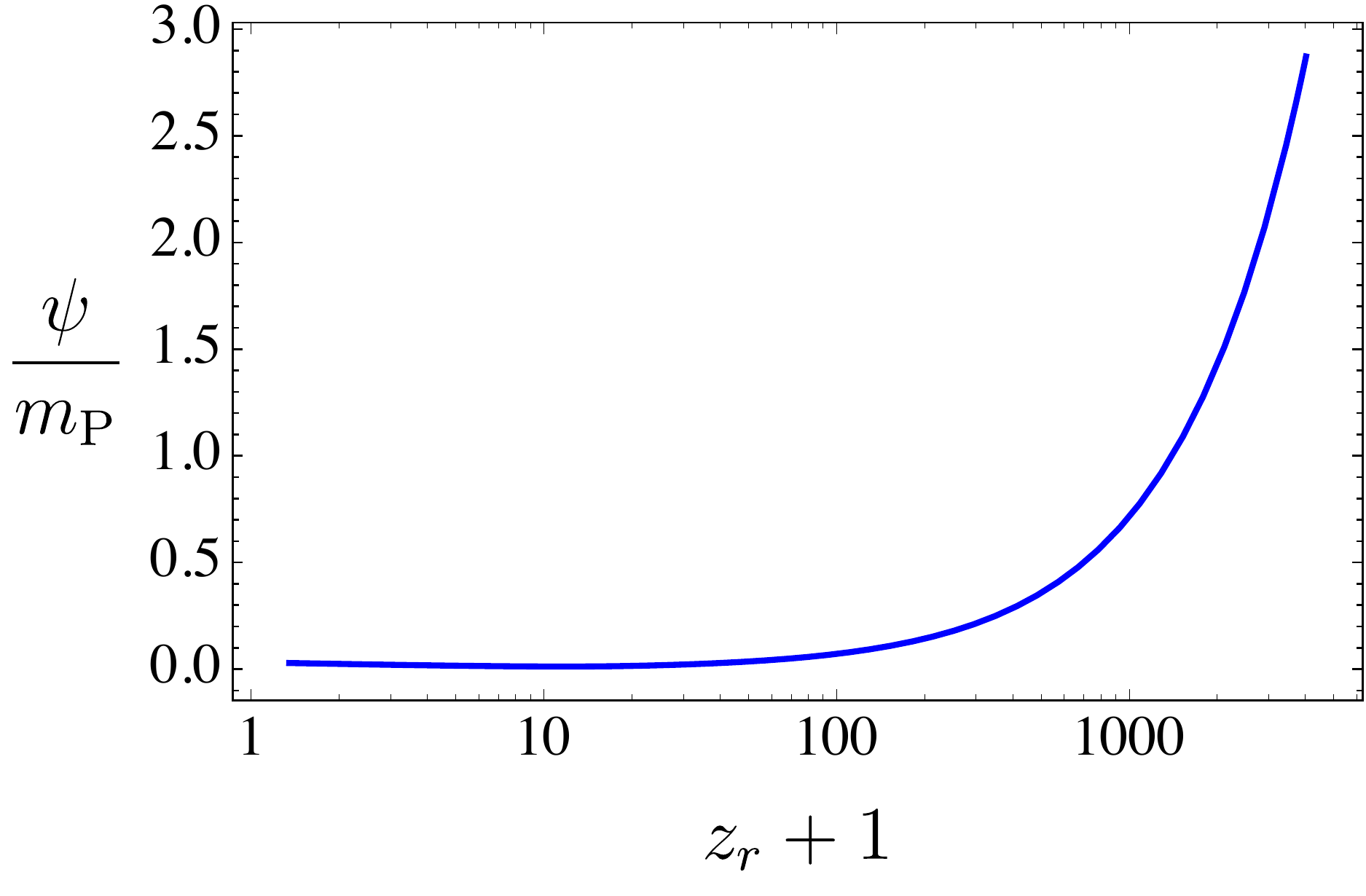}
}
\vspace{0.05cm}

{
  \centering
  \includegraphics[width=0.98\columnwidth]{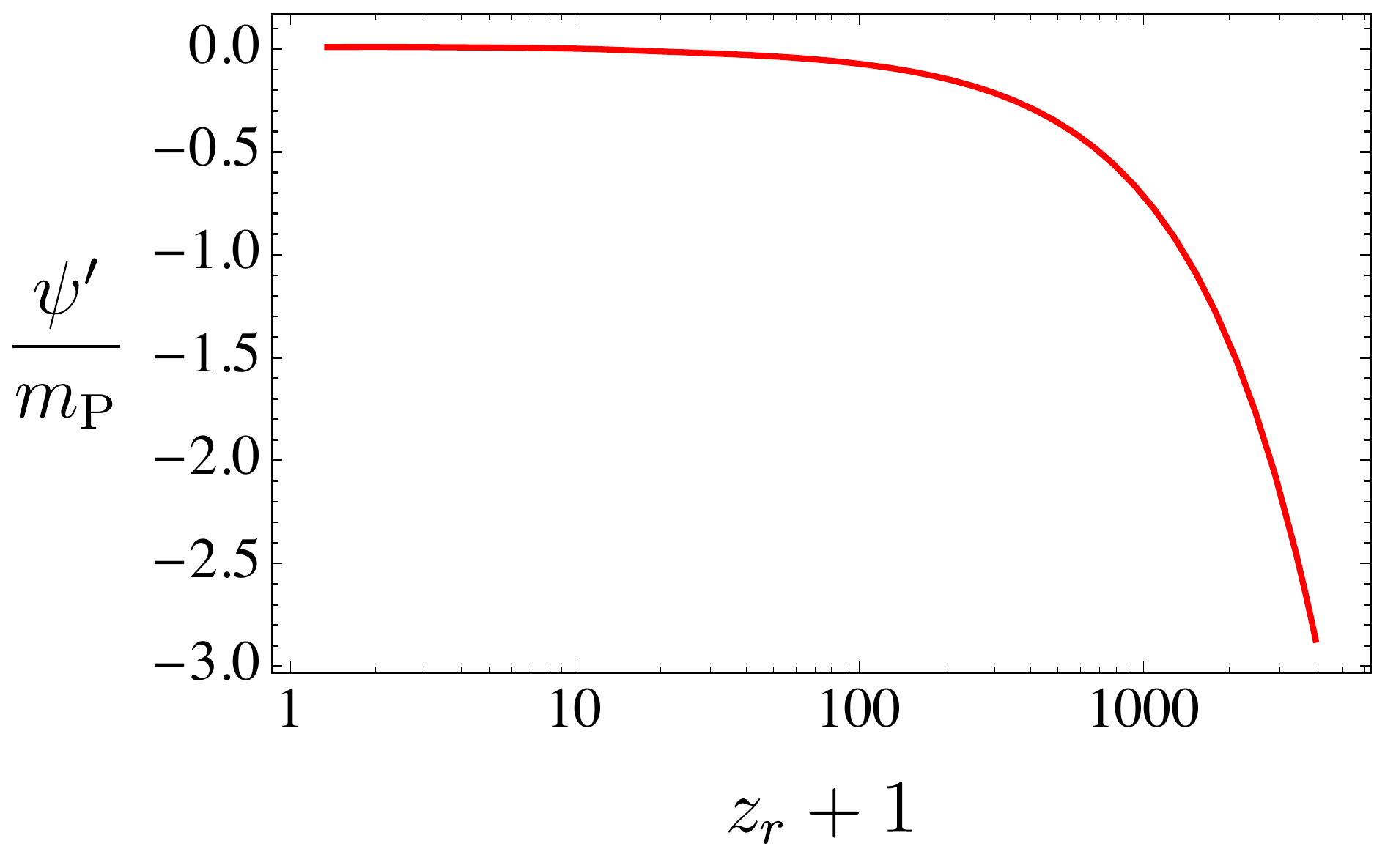}
}
\caption{Evolution of the gauge field (upper), and its speed (lower), during the matter dominated period. The gauge field is still decaying in a decelerated way (in magnitude).}
\label{Matter}
\end{figure}

This period runs from $z_r \simeq 4000$ to $z_r \simeq  0.3$. In Fig. \ref{Matter}, we can see that during this epoch the gauge field is still decaying in a decelerated way (in magnitude). The contribution of dark energy around $z_r = 50$ is $\Omega_{\text{DE}} \approx 2 \times 10^{-5}$, well below the CMB constraint $\Omega_{\text{DE}} < 0.02$ \cite{Ade:2015rim}. During this period, the main contribution to the dark sector comes from the Yang- Mills term, however, it is possible to find initial conditions where the mass term is the main dark component.

%%%%%%%%%%%%%%%%%%%%%%%%%%%%%%%%%%%%

\subsubsection{Dark energy dominated period}

\begin{figure}[t!]
{
\centering
\includegraphics[width=0.98\columnwidth]{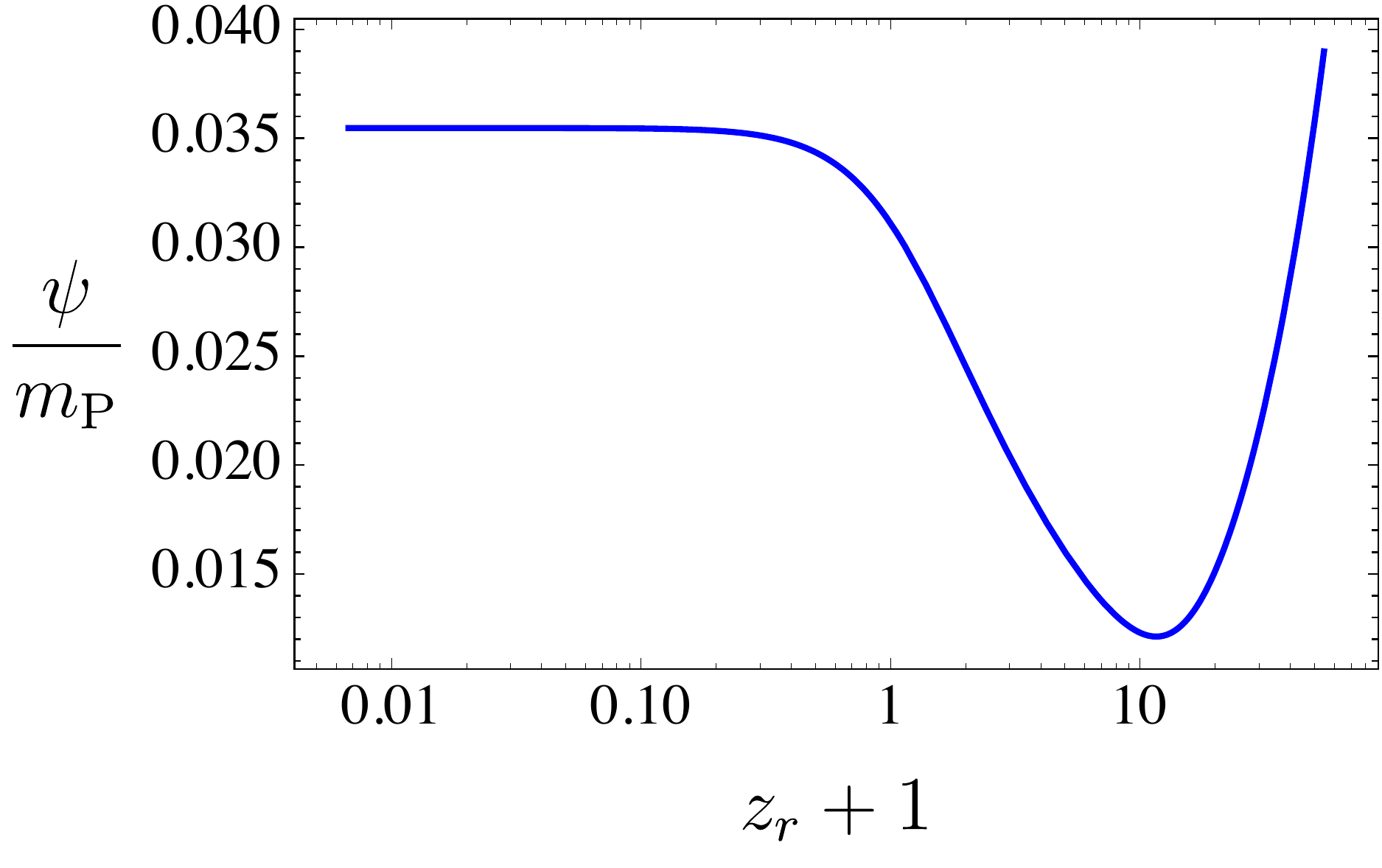}
}
\vspace{0.05cm}

{
  \centering
  \includegraphics[width=0.98\columnwidth]{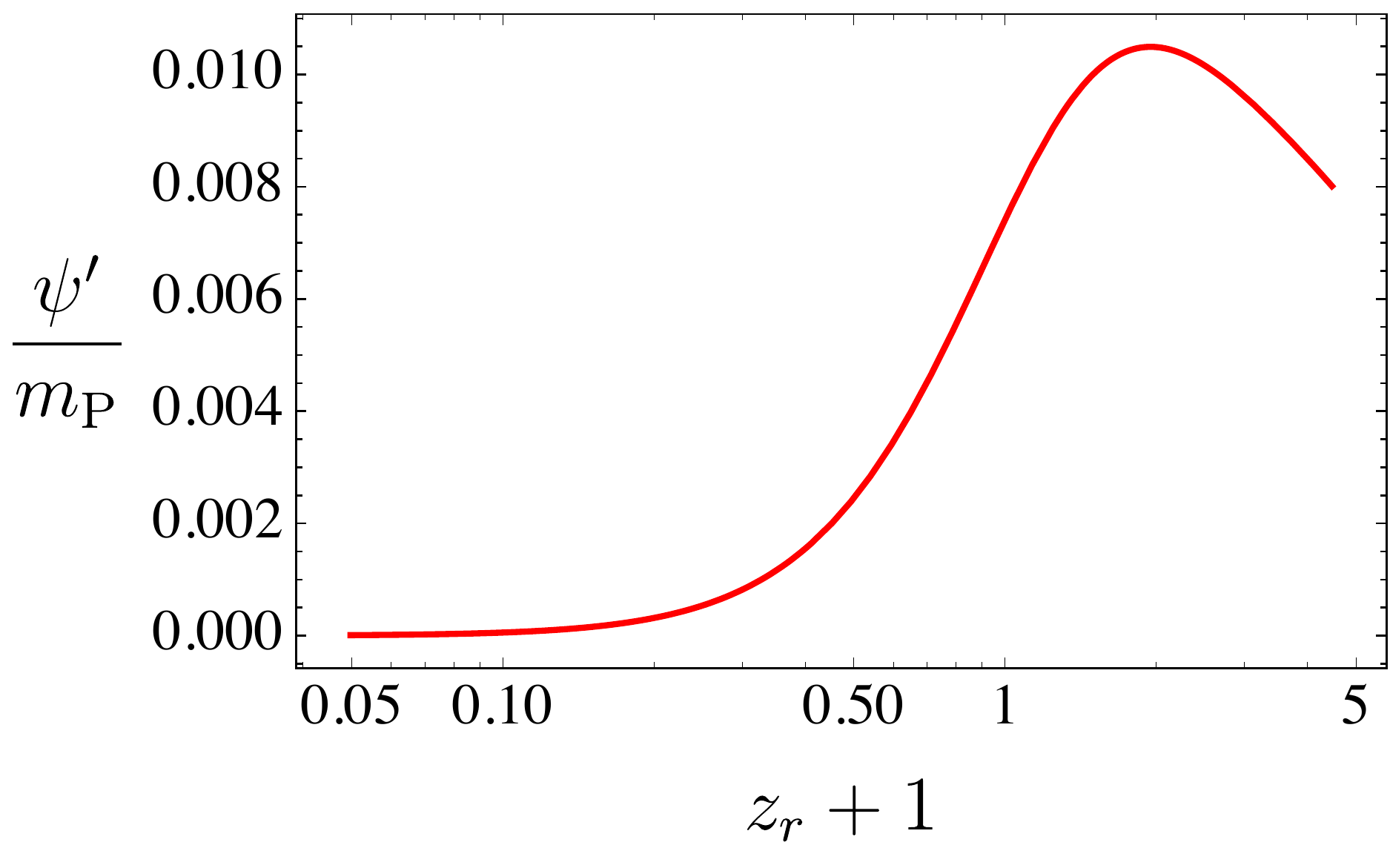}
}
\caption{Evolution of the gauge field (upper), and its speed (lower), during the dark energy dominated period. The gauge field $\psi$ reaches a constant value at late times, where $\dot{\psi} = 0$, and it becomes an effectively cosmological constant.}
\label{Dark}
\end{figure}

This period runs from $z_r \approx 0.3$ on into the future. During this epoch (Fig. \ref{Dark}), the gauge field decays until it reaches a minimum and after that a constant value, when its speed goes to zero, as expected since $x = z$ in the point (\emph{DE}) (dark energy dominance). The asymptotic value of the gauge field is $ \psi \approx 0.0354378 \, m_\text{P} \,,$ in agreement with $\sqrt{2} \, z = 0.0354917$ which is the value in the attractor point (\emph{DE}). 

We also can perform a rough estimation of the parameters $\kappa$ and $g$. Replacing $y_0$ and $z_0$ in Eq.  (\ref{GF H}) we get\footnote{The subscript $0$ means that the corresponding quantity is evaluated today.}
\begin{equation}
\frac{H_0}{g} = \frac{\sqrt{2} z_0^2}{y_0} m_\text{P} \approx 0.042 \, m_\text{P} \,,
\end{equation}
and using the observational value $H_0 \approx 10^{- 61} m_\text{P}$ \cite{Ade:2015xua} we find
\begin{equation}
g \approx 2.38 \times 10^{-60} \,,\, \,  \kappa \approx 1.76 \times 10^{130} m_\text{P}^{-4} \,.
\end{equation}

This calculation shows that the gauge coupling $g$ is extremely small (the order of $10^{-60}$), while the parameter $\kappa$ is extremely large  (the order of $10^{130} m_\text{P}^{-4}$). 

%%%%%%%%%%%%%%%%%%%%%%%%%%%%%%%%%%%%

\subsection{Dark energy equation of state} \label{Eq of state}

From the above numerical solution we can say that, at late times, the dark sector behaves very similar to a cosmological constant once the $\kappa$-term is dominating the energy budget. However, in order to thoroughly characterize the behaviour of this sector, it is necessary to study the evolution of its equation of state. 

\begin{figure}[h!]
\centering
\includegraphics[width=\linewidth]{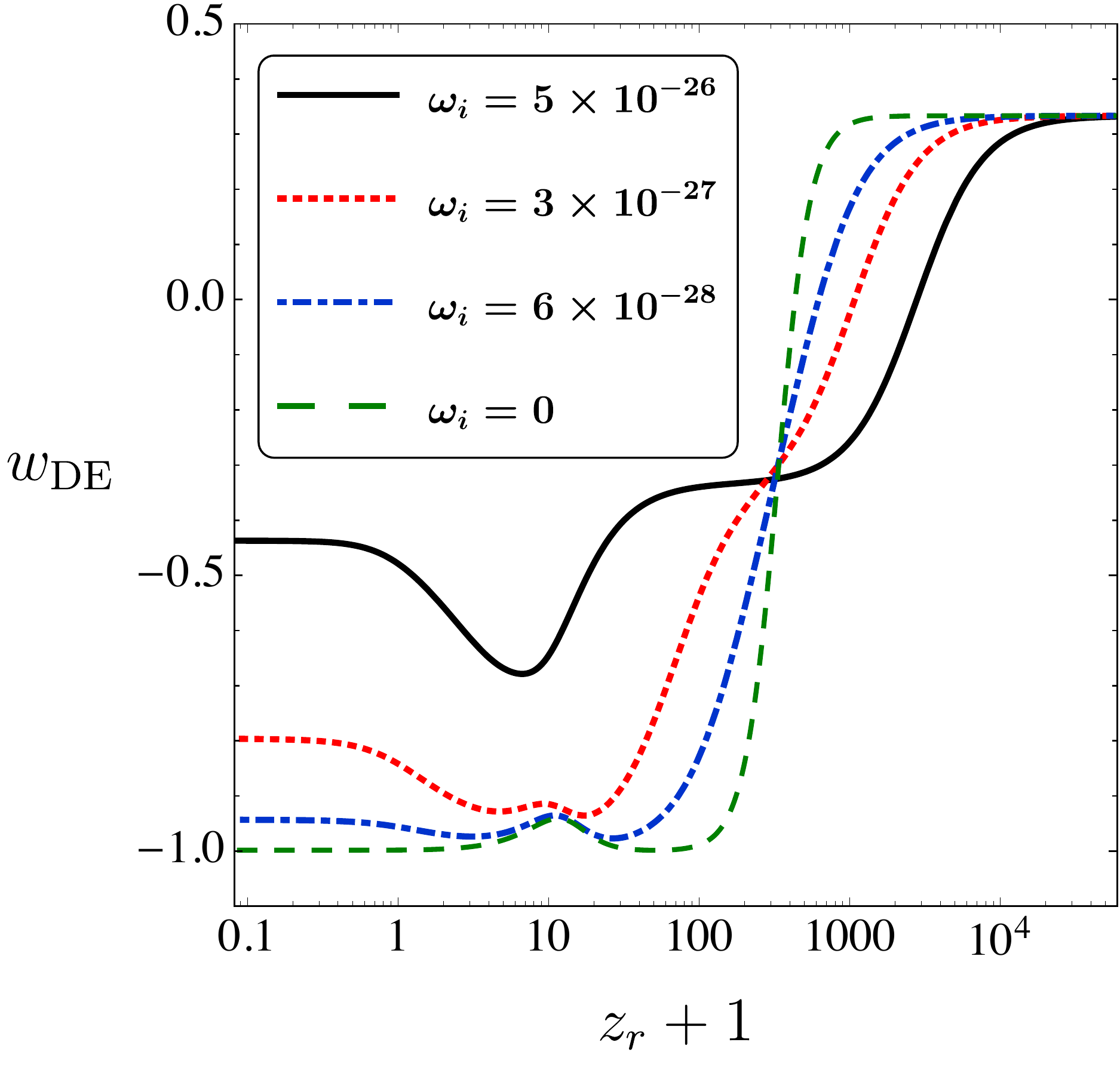}
\caption{Evolution of $w_\text{DE}$ for different $\omega$. Arbitrary values of $\omega$ are not viable since they do not yield to a correct expansion history of the Universe. The initial state corresponds to ``dark radiation", and the final stage to dark energy. The values of $\alpha, x, z, y$ and $\Omega_{r}$ are the same used in the numerical analysis section.}
\label{DEEoS}
\end{figure}

Because of the Friedmann constraint in Eq. (\ref{de constraint}), the values of the variables $w_i$ and $y_i$ are determined by $\omega_i$ (see Eq. (\ref{de omega})). In Fig. \ref{DEEoS} we plot $w_\text{DE}$ for several values of $\omega_i$. For $z_r > 10^4$ we see that $w_\text{DE} \approx 1 / 3 $, meaning that the dark sector behaves as a radiation fluid at early times. This is expected since the Yang-Mills term is dominating over the dark components during this period, so this term can contribute to the early relativistic degrees of freedom as claimed in Ref. \cite{Mehrabi:2017xga}. We also see that for $\omega_i \neq 0$ it is possible that the mass term dominates the dark sector implying $w_\text{DE} \approx - 1 / 3$. This is a new behaviour with respect to the results given in Ref. \cite{Mehrabi:2017xga}. As a comment, a fluid with equation of state equal to $- 1 / 3$ cannot drive accelerated expansion. However, there is no  transition period from decelerated expansion to accelerated expansion in our model, because dust is the dominant fluid when $w_\text{DE} \approx - 1 / 3$ and thus $w_{\text{eff}} \approx 0$. Note that in all of the cases, the final stage of the Universe is dark energy domination, which is reached at different cosmological epochs, depending on the value of $\omega_i$. This behaviour takes place due to modifications in the Hubble parameter since a change in $\omega$ implies a change in $H / g$ by the relation in Eq. (\ref{GF H}), once the other variables are fixed.
 
In Fig. \ref{DEEoS} we realized that the point (\emph{S}), i.e. the matter-dark energy scaling solution, is irrelevant to the cosmological dynamics. This can be noticed since in this point $w_\text{DE} = 0$ but the equation of state of dark energy never spends much time in this stage, so (\emph{S}) is not a metastable point as  (\emph{R}) and  (\emph{M}) are. In this plot, we also see that $w_\text{DE}$ has some peaks around $z_r = 10$. This is because, around this time, the gauge field reaches its minimum magnitude, as can be seen in Fig. \ref{Dark}. We want to stress that this particular behaviour of $w_{\text{DE}}$ is a distinctive property found in this model that was not reported in Ref. \cite{Mehrabi:2017xga}.

As expected $-1 < w_\text{DE} < 1/3$, contrasting with the equation of state of any quintessence model where $ - 1 < w_\text{DE} < 1$. The main difference is in the so-called ``kination" period, which corresponds to the epoch when the kinetic term of the quintessence field dominates over its potential. This period must be prior to the radiation  dominance epoch since the energy density of the quintessence field decays as $a^{-6}$ \cite{Amendola:2015ksp}. As a comment,  some authors claim that this kination period could be very useful in the study of the reheating process (see e.g. Refs. \cite{Visinelli:2017qga,Bettoni:2018utf,Bettoni:2018pbl,Dimopoulos:2018wfg,Bettoni:2019dcw,Opferkuch:2019zbd}).

%%%%%%%%%%%%%%%%%%%%%%%%%%%%%%%%%%%%

\subsection{Anisotropic massless case}

So far, we have discussed systems embedded in an FLRW metric. It was shown that any early background spatial shear is extremely damped within a few $e$-folds during the inflationary phase \cite{Maleknejad:2011jr}. In contrast, the vector gauge field acquires a constant magnitude in the late-time cosmological expansion. This suggests that the gauge field could support an anisotropic expansion. In the following, we investigate this possibility where, by simplicity, we only consider massless gauge fields. 

We assume a homogeneous but anisotropic spacetime described by the axially symmetric Bianchi-I metric
\begin{equation}
\text{d} s^2 = - \text{d} t^2 + e^{2 \alpha}\left[ e^{-4\sigma} \text{d} x^2 + e^{2\sigma} \left( \text{d} y^2 + \text{d} z^2 \right)\right] \,,
\end{equation}
where $e^{\alpha(t)} = a(t)$ is the average scale factor and $\sigma(t)$ is the shear.

An appropriate axially symmetric ansatz for the gauge field is
\begin{equation} \label{axial ansatz}
A^a_{\,\, 0}(t) = 0 \,, \quad A^a_{\,\, i}(t) = e^a_{\ i}(t) \psi_i(t) \,,
\end{equation}
where
\begin{equation}
e^1_{\ 1}(t) = e^{\alpha(t) - 2 \sigma(t)} \,,\quad e^2_{\ 2}(t) = e^3_{\ 3}(t) = e^{\alpha(t) + \sigma(t)} \,,
\end{equation}
\begin{equation}
\psi_1(t) = \frac{\psi(t)}{\lambda^2(t)} \,,\quad \psi_2(t) = \psi_3(t) = \lambda(t) \psi(t) \,,
\end{equation}
with $\lambda(t)$ a function parametrizing the deviation from the isotropic configuration of the gauge field. This kind of ansatz has been used in other works where the dynamics of gauge fields in anisotropic backgrounds is studied (see Refs. \cite{Maleknejad:2011jr, Maleknejad:2013npa, Murata:2011wv}).

Employing the axial ansatz in the energy tensor in Eq. (\ref{energy tensor}), and assuming massless gauge vector fields, i.e. $m_a = 0$, the density and pressure coming from the Yang-Mills term are
\begin{multline}
\rho_{\text{YM}} = \frac{1}{2 \lambda^4} \left[ \frac{\dot{\phi}}{a} - 2 \frac{\phi}{a} \left( \dot{\sigma} + \frac{\dot{\lambda}}{\lambda} \right)\right]^2  \\
+ \lambda^2 \left[ \frac{\dot{\phi}}{a} + \frac{\phi}{a} \left( \dot{\sigma} + \frac{\dot{\lambda}}{\lambda} \right)\right]^2 + \frac{g^2 \phi^4}{2 a^4} \frac{2 + \lambda^6}{\lambda^2} \,,
\end{multline}
and $p_{\text{YM}} = \rho_{\text{YM}}/3$. For the $\kappa$-term, we have $p_\kappa = - \rho_\kappa$ with $\rho_\kappa$ given by Eq. (\ref{densities}) meaning that this term does not introduce anisotropies in the energy tensor.

The corresponding Friedmann equations are
\begin{equation}
3 m_\text{P}^2( H^2 - \dot{\sigma}^2) = \rho_{\text{YM}} + \rho_\kappa + \rho_r + \rho_m  \,,
\end{equation}
\begin{equation}
 m_\text{P}^2 (\dot{H} + 3 \dot{\sigma}^2) = - \left[ \frac{2}{3} \rho_{\text{YM}} + \frac{2}{3} \rho_r + \frac{1}{2}\rho_m  \right]\,,
\end{equation}
\begin{multline}
 m_\text{P}^2 (\ddot{\sigma} + 3 H \dot{\sigma}) = \frac{1}{3 \lambda^4} \left[ \frac{\dot{\phi}}{a} - 2 \frac{\phi}{a} \left(\dot{\sigma} + \frac{\dot{\lambda}}{\lambda} \right) \right]^2 \\- \frac{g^2 \phi^4}{3 a^4} \frac{1 - \lambda^6}{\lambda^2}   
- \frac{\lambda^2}{3} \left[ \frac{\dot{\phi}}{a} + \frac{\phi}{a} \left(\dot{\sigma} + \frac{\dot{\lambda}}{\lambda} \right) \right]^2 \,.
\end{multline}
The axial ansatz in the equations of motion for the gauge field components yields to two related equations given by
\begin{multline}
0 = \frac{\ddot{\phi}}{a} \left[ 1 + \frac{1}{3}\kappa g^2 \frac{\phi^4}{a^4} \frac{2 + \lambda^6}{\lambda^2}\right] + H \frac{\dot{\phi}}{a} \left[ 1- \kappa g^2 \frac{\phi^4}{a^4} \frac{2 + \lambda^6}{\lambda^2} \right] \\[1mm]
 - 2 \frac{\phi}{a} \left( \dot{\sigma}^2 - \frac{\dot{\lambda}^2}{\lambda^2} \right) + \frac{2}{3} \frac{g^2 \phi^3}{a^3} \frac{1 + 2 \lambda^6}{\lambda^4}   \\[1mm]
+ \frac{2}{3} \kappa g^2 \frac{\phi^3 \dot{\phi}^2}{a^5} \frac{2 + \lambda^6}{\lambda^2} \,,
\end{multline}
and 
\begin{multline}
0 = \frac{\ddot{\lambda}}{\lambda} + 2 \frac{\dot{\phi}}{\phi} \frac{\dot{\lambda}}{\lambda} + \ddot{\sigma} - \dot{\sigma}^2  + H \left( \dot{\sigma} + \frac{\dot{\lambda}}{\lambda} \right) + \frac{g^2 \phi^2}{a^2} \frac{1 + \lambda^6}{\lambda^4}   \\
+ \frac{\ddot{\phi}}{\phi}\left( 1 + \kappa g^2 \frac{\phi^4}{\lambda^2 a^4} \right) + \frac{H \dot{\phi}}{\phi} \left(1 - 3\kappa g^2 \frac{\phi^4}{\lambda^2 a^4} \right) \,.
\end{multline}

\begin{figure}[h!]
{
\centering
\includegraphics[width=\columnwidth]{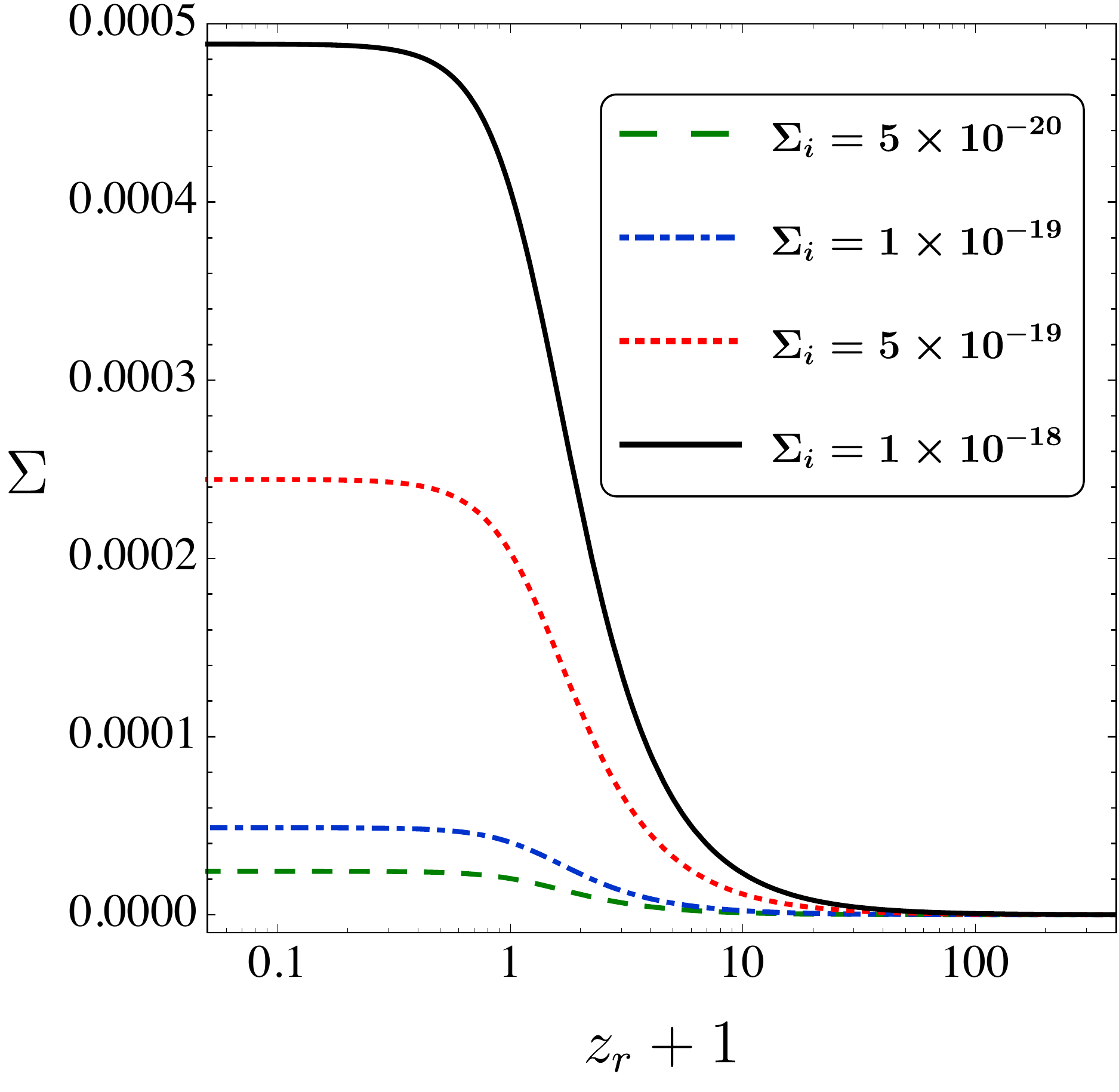}
}
\vspace{0.05cm}

{
  \centering
  \includegraphics[width=\columnwidth]{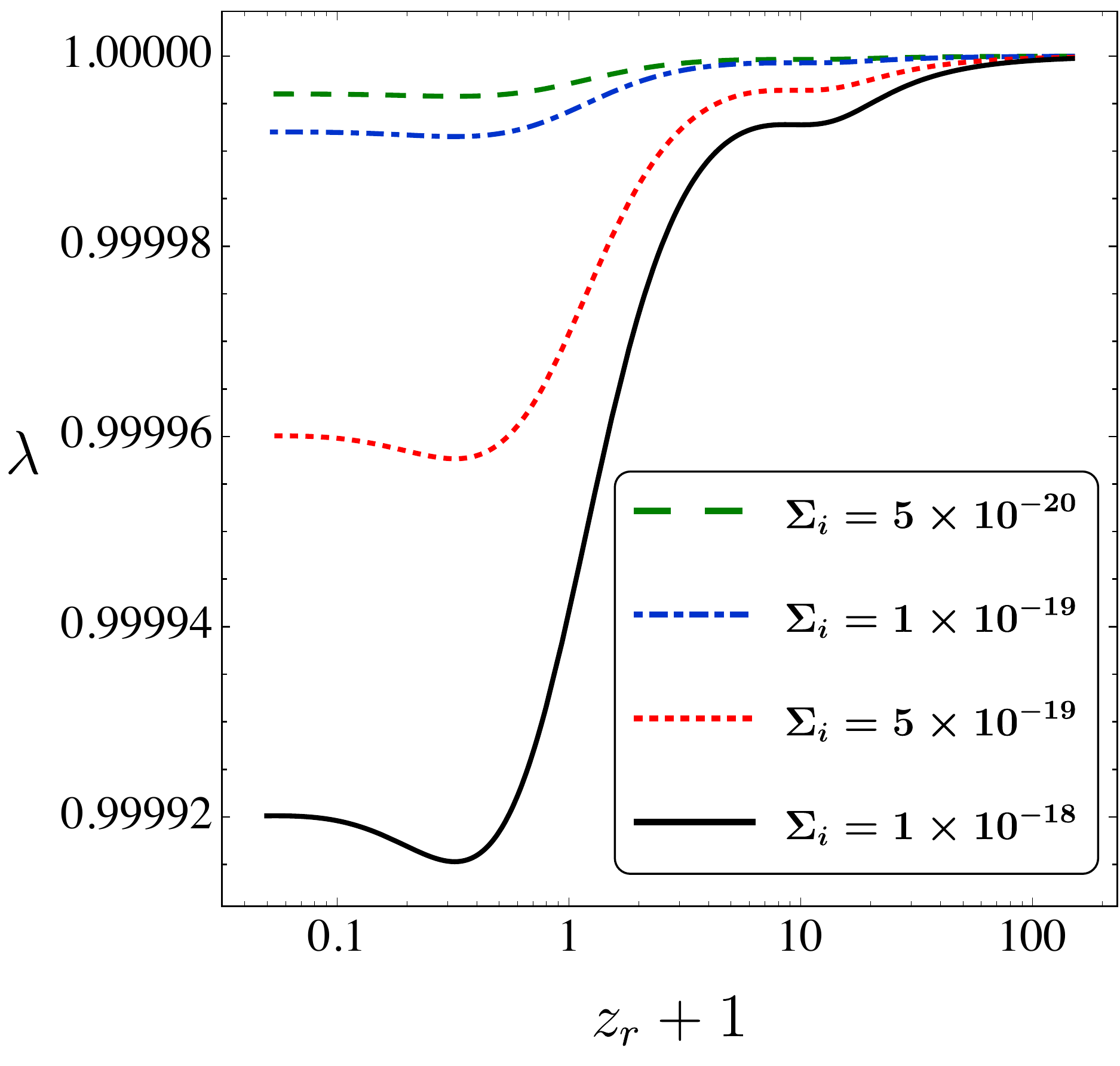}
}
\caption{Evolution at late times of the anisotropic degrees of freedom. The initial conditions were chosen in the deep radiation epoch, corresponding to the same ones used in the isotropic model with $l = 1-10^{-20}$ and $s = 10^{-20}$, while $\Sigma$ varies. Even a small deviation from the isotropic initial conditions yields to a non-negligible anisotropic contribution to the present Universe density budget.}
\label{Anisotropies}
\end{figure}

The value of the present shear is constrained to be $| \Sigma_0 | \leq \mathcal{O}(0.001)$ \cite{Campanelli:2010zx, Amirhashchi:2018nxl}. More restricted bounds are expected from future observational missions like Euclid \cite{Amendola:2016saw}. Since observations rule out high anisotropies, we are interested in anisotropic solutions near to the isotropic solutions obtained in the last section. Therefore, we rewrite the system using the same expansion variables defined in Eqs. (\ref{inf variables}) and (\ref{de variables}). The dynamical degrees of freedom involving the anisotropy in the gauge field and the background are encoded in the variables
\begin{equation}
l \equiv \lambda^2 \,,\quad s \equiv \frac{\dot{\lambda}}{\lambda \, H} \,,\quad \Sigma \equiv \frac{\dot{\sigma}}{H} \,,
\end{equation}
such that, the isotropic limit corresponds to $l = 1,\, s = 0$ and $\Sigma = 0$. In this case, the dynamical systems technique provides an autonomous set very hard to deal with. We present the full system in Appendix \ref{AppendixA}. Instead of looking for the fixed points of the whole system, we numerically integrate the system around the isotropic solutions found in the last section. Explicitly, we use the same initial conditions of the isotropic case, with $\omega_i = 0$, and assume that
\begin{equation}
l_i = 1 - 10^{-20} \,,\quad s_i = 10^{-20} \,,
\end{equation}
at $z_r = 2.18 \times 10^8$, while $\Sigma_i$ varies two orders of magnitude from a very small value. As seen in Fig. \ref{Anisotropies}, even a small deviation from the initial conditions used in the isotropic model yields a non-negligible amount of anisotropy in the present Universe. The values obtained are within the observational bounds, explicitly we found $| \Sigma_0 | < 5 \times 10^{-4}$.

The equation of state of dark energy is also modified by the anisotropy in the following way:
\begin{equation}
w_{\text{DE}} = \frac{p_{\text{DE}}}{\rho_{\text{DE}}} = \frac{1}{3} \frac{\Omega_{\text{YM}} - 3 \Omega_\kappa + 3 \Sigma^2}{\Omega_{\text{YM}} + \Omega_\kappa + \Sigma^2} \,,
\end{equation}
where $\Omega_{\text{YM}} \equiv \rho_{\text{YM}} / 3 m_{\text{P}}^2 H^2$ and $\Omega_\kappa \equiv \rho_\kappa / 3 m_{\text{P}}^2 H^2$ are the density parameters for the Yang-Mills and $\kappa$-terms, respectively. However, since $\Sigma^2 \ll \Omega_\kappa, \Omega_{\text{YM}}$ during the whole expansion history, the contribution of the shear is always negligible in comparison to the other dark components and thus the changes are not noticeable. We want to stress that, although the asymptotic behaviour of the model remains unknown, this model could support a late-time anisotropic expansion observable nowadays. It is possible that the Universe lose its anisotropic hair in the future, as it is the case presented e.g. in Ref. \cite{Orjuela-Quintana:2020klr}. We left a better exploration of the cosmological consequences of this model for a future work.

%%%%%%%%%%%%%%%%%%%%%%%%%%%%%%%%%%%%

\section{Conclusions} \label{conclusions}

In this paper, we studied non-Abelian gauge vector fields endowed with SU(2) group representation as the unique source of inflation and dark energy. In the inflationary scenario, it was shown that this primordial accelerated expansion can be driven solely by this kind of fields \cite{Maleknejad:2011sq}. Since this model, known as gaugeflation, was ruled out by observations \cite{Namba:2013kia}, several modifications to the original model have been proposed. In particular, the introduction of a mass term could ameliorate the tension \cite{Nieto:2016gnp, Adshead:2017hnc}. However, in these previous works, only some particular inflationary trajectories were analyzed. Here, by using a dynamical system approach, we have extended their results, finding the full available parameter space yielding to slow-roll inflationary solutions. We realized that slow-roll inflation is a saddle point in the phase space and thus this model provides a natural mechanism to end the inflationary period. We also found that the inclusion of the mass term increases the length of the slow-roll phase, instead of reducing it as claimed in Ref. \cite{Nieto:2016gnp}. This reduction is because the authors of Ref. \cite{Nieto:2016gnp} ignored the relation between the parameters $\gamma_i$ and $\omega_i$ given by the Friedmann constraint, once the other initial conditions have been set. Hence an increase in $\omega_i$ implied an increase in the speed of the gauge field $\dot{\psi}_i$ yielding to a drastic reduction in the number of $e$-folds. Thereafter, we remark that our conclusion could modify the results obtained in Ref. \cite{Adshead:2017hnc} where the linear treatment of the model was worked out. For example, the relation between $\gamma_i$ and $\omega_i$ changes the results of $n_\text{scal}/\epsilon$ (see Fig. \ref{nscal}). To respond to this query, we plan to make a full treatment of perturbations for  the massive case in a future work.

Regarding the late-time accelerated expansion, we were able to show that the dark energy domination period is the only physical attractor of the theory, and thus confirming the results in Ref. \cite{Mehrabi:2017xga} where only some particular sets of initial conditions and parameters were studied. We also generalized the model of Ref. \cite{Mehrabi:2017xga} by introducing a mass term to the dynamics. We found that this term yields to new behaviors in the equation of state of dark energy $w_\text{DE}$. For instance, some peculiar peaks around the redshift $z_r = 10$ were observed (see Fig. \ref{DEEoS}). By noting that the gauge field acquires a constant magnitude in the attractor point, we studied the same action but in a homogeneous and anisotropic axially symmetrical Bianchi-I background. The dynamical system of the model is very cumbersome (see Appendix \ref{AppendixA}) hence we opted for numerical integration of the autonomous set. Since observations rule out high anisotropies \cite{Campanelli:2010zx, Amirhashchi:2018nxl} we set the initial conditions near to the isotropic solutions. We found that, starting from a high isotropic Universe, the gauge field can support a late-time anisotropic expansion, given that the spatial shear evaluated today is within the observational bounds. A more rigorous treatment on anisotropic models of dark energy driven for non-Abelian gauge fields is left for future works. 

%%%%%%%%%%%%%%%%%%%%%%%%%%%%%%%%%%%%

\section*{ACKNOWLEDGEMENTS}
This work was supported by the following grants: Vicerrector\'ia de Investigaciones $-$ Universidad del Valle Grant No. 71220, and Vicerrector\'ia de Ciencia, Tecnología, e Innovación $-$ Universidad Antonio Nari\~no Grants  No. 2019248 and No. 2019101. The authors want to thank Yeinzon Rodr\'iguez and Carlos Nieto for useful discussions.

%%%%%%%%%%%%%%%%%%%%%%%%%%%%%%%%%%%%

\appendix

%%%%%%%%%%%%%%%%%%%%%%%%%%%%%%%%%%%%%%%
\section{Effect of the Mass Term on the Length of Inflation} \label{N vs m}
%%%%%%%%%%%%%%%%%%%%%%%%%%%%%%%%%%%%%%%

In Sec. \ref{Massive GF}, we studied the effect that the mass term has on the length of the inflationary period, showing that an increase in the mass implies an increase in the expected number of $e$-folds. We reached that conclusion after analyzing the behaviour of $N$ with respect to the mass parameter $\omega$ (see Fig. \ref{MGF efolds plot}). Here, we write $N$ explicitly in terms of $m$. 

By fixing the parameters ($g$ and $\kappa$) and the initial values for the field ($\psi_i$ and $\dot{\psi}_i$), the Friedmann equation (\ref{inf H2}) becomes a quadratic equation for the Hubble parameter, $H_i$, in terms of $m$
\begin{align} \label{Quad H}
3 m_\text{P}^2 H_i^2 &= \frac{3}{2} \Big[ \left( \psi_i H_i + \dot{\psi}_i \right)^2 + g^2 \psi_i^4 \nonumber \\
 &+ \kappa g^2 \psi_i^4 \left( \psi_i H_i + \dot{\psi}_i \right)^2 + m^2 \psi_i^2 \Big]
\end{align}
Having $H_i$ in terms of $m$, i.e. $H_i = H_i (m)$, we can write the parameters of the model in terms solely of $m$. The slow-roll parameter in Eq. (\ref{inf slowroll}) is
\begin{equation}
\epsilon_i (m) = \left( \psi_i + \frac{\dot{\psi}_i}{H_i (m)} \right)^2 + \frac{g^2 \psi_i^4}{H_i^2 (m)} + \frac{m^2 \psi_i^2}{2 H_i^2 (m)} \,,
\end{equation}
while the parameters $\gamma_i$ and $\omega_i$ in Eqs. (\ref{inf gamma}) and (\ref{inf omega}) are
\begin{equation}
\gamma_i (m) = \frac{g^2 \psi_i^2}{H_i^2 (m)} \,,\quad \omega_i (m) = \frac{m^2}{H_i^2 (m)} \,. 
\end{equation}
Therefore, the approximated number of $e$-folds,
\begin{equation}
N (m) \approx \frac{1 + \gamma_i + \omega_i/2}{2 \epsilon_i} \,\, \text{ln} \left( \frac{1 + \gamma_i + \omega_i/2}{\gamma_i + \omega_i/2} \right) \,,
\end{equation}
can be expressed as a function of the mass, and thus the effect of $m$ on the length of inflation can be isolated.

\begin{figure}
\centering
\includegraphics[height = 5.2cm, width=0.9\linewidth]{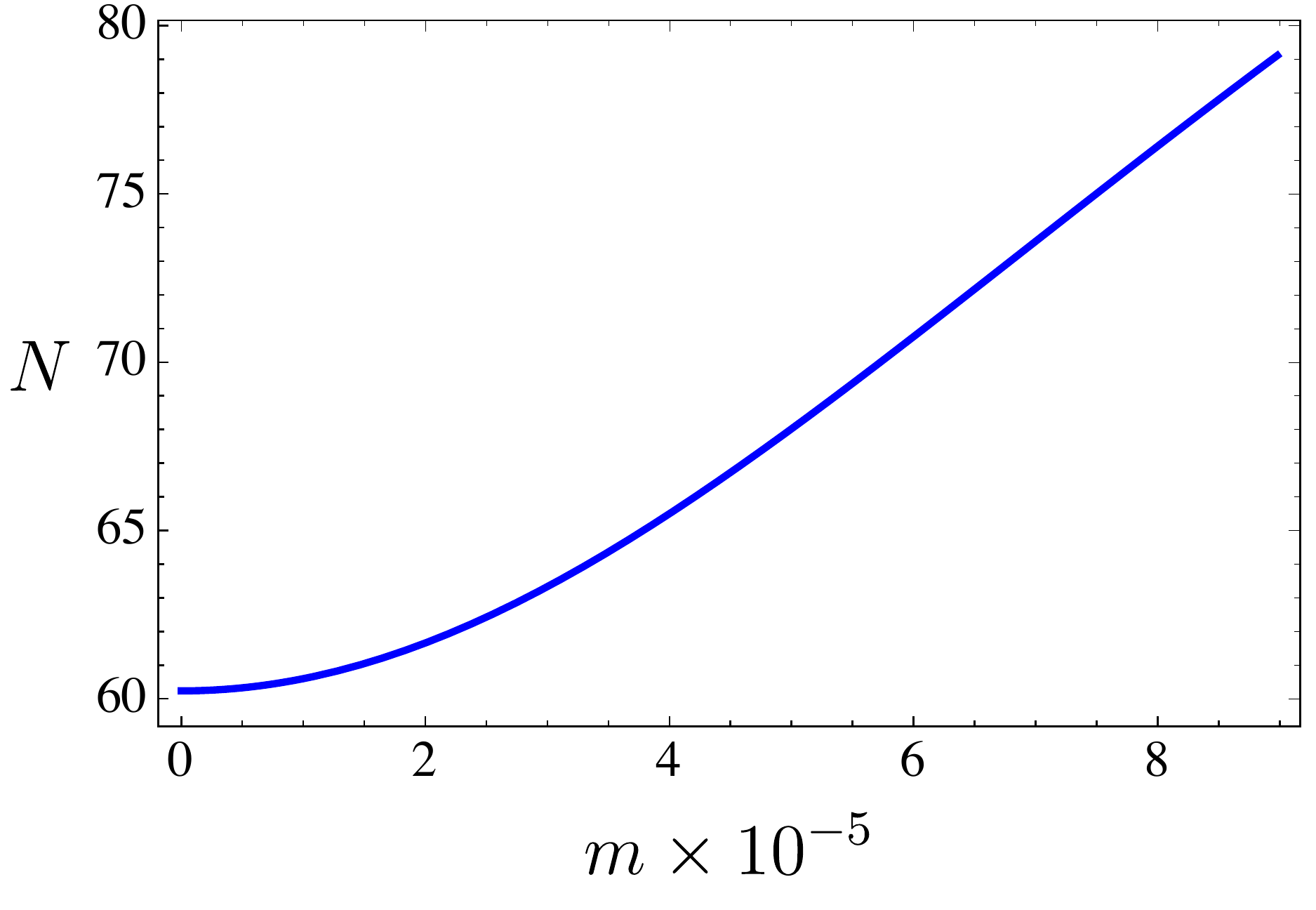}
\caption{Number of $e$-folds in function of $m$. The mass term increases the length of the inflationary phase.}
\label{Nvsm}
\end{figure}

As an example, let us assume that
\begin{equation*}
g = 2.5 \times 10^{-3} \,,\quad \kappa = 1.733 \times 10^{14} \,,
\end{equation*}
\begin{equation}
\psi_i = 0.035 \,,\quad \dot{\psi}_i = 0 \,.
\end{equation}
For this particular set of parameters and initial conditions, the solution of the quadratic equation (\ref{Quad H}) gives 
\begin{equation}
H_i \approx \sqrt{1.2 \times 10^{-9} \, m_\text{P}^2 + 0.16 \, m^2} \,,
\end{equation}
Supposing that the energy scale of inflation is $H_i < 5 \times 10^{-5} m_\text{P}$, the mass is bounded to be
\begin{equation}
0 \leq m \lesssim 9 \times 10^{-5} m_\text{P} \,.
\end{equation}
In Fig. \ref{Nvsm}, we plot the expected number of $e$-folds in terms of the mass of the field. From this figure, it is clear the effect that the mass term has on the model: it increases the length of the inflationary phase.

%%%%%%%%%%%%%%%%%%%%%%%%%%%%%%%%%%%%%%%
\section{Anisotropic Dark Energy: Autonomous System} \label{AppendixA}
%%%%%%%%%%%%%%%%%%%%%%%%%%%%%%%%%%%%%%%

We present the full autonomous system obtained from the expansion variables $x, y, z, \Omega_r, \Sigma, l$ and $s$.

\begin{align}
x' &= x q + p \,,\\[1mm]
y' &= y \left( 2 \frac{x}{z} + q -1 \right) \,,\\[1mm]
z' &= x - z \,,\\[1mm]
\Omega'_r &= 2 \Omega_r \left(q - 1\right) \,,\\[1mm]
\Sigma' &= \Sigma (q + 1) + u \,,\\[1mm]
l' &= 2 l \, s \,, \\[1mm]
s' &= s (q + 1) - s^2 + v \,,
\end{align}
where the deceleration parameter has the form
\begin{align}
q = \frac{1}{2} &\Big( 1 + \frac{1}{3 l^2} \left[ x - 2z (\Sigma + s) \right]^2 + y^2 \frac{2 + l^3}{3 l} + 3 \Sigma^2 \nonumber \\ 
 &+ \frac{2}{3}l \left[ x + z (\Sigma + s) \right]^2 -12 \alpha x^2 z^4 + \Omega_r \Big)
\end{align}
and the functions 
\begin{equation}
p \equiv \frac{1}{\sqrt{2} M_\text{pl}}\frac{\ddot{\phi}}{a H^2} \,,\quad u \equiv \frac{\ddot{\sigma}}{H^2} \,,\quad v \equiv \frac{\ddot{\lambda}}{\lambda H^2} 
\end{equation}
obey the equations 
\begin{align}
0 &= p \left[ 1 + \frac{4}{3}\alpha z^4 \frac{2 + l^3}{l} \right] + x \left[ 1 - 4 \alpha z^4 \frac{2 + l^3}{l} \right] \nonumber \\
 &+\frac{8}{3} \alpha x^2 z^3 \frac{2 + l^3}{l} + \frac{2}{3} \frac{y^2}{z} \frac{1+ 2 l^3}{l^2} - 2z \left( \Sigma^2 - s^2 \right) \,,
\end{align}
\begin{align}
u &= \frac{2}{3 l^2} \left[ x - 2z (\Sigma + s) \right]^2 - \frac{2}{3} l \left[ x + z (\Sigma + s) \right]^2 \nonumber \\
 &- \frac{2}{3} y^2 \frac{1 - l^3}{l} - 3 \Sigma \,,
\end{align}
and
\begin{align}
0 &= v + 2 \frac{x}{z} s + u - \Sigma^2 + \Sigma + s + \frac{p}{z} \left( 1 + 4\alpha \frac{z^4}{l} \right) \nonumber \\
 &+ \frac{x}{z} \left( 1 - 12 \alpha \frac{z^4}{l} \right) + \frac{y^2}{z^2} \frac{1 + l^3}{l^2} + 8\alpha \frac{x^2 z^3}{l} \,.
\end{align}

%%%%%%%%%%%%%%%%%%%%%%%%%%%%%%%%%%%%

\bibliographystyle{utcaps} 
\bibliography{Bibli.bib} 

\end{document}